\documentclass[aps,prl,reprint,superscriptaddress,nopacs]{revtex4-1}

\usepackage{graphicx}% Include figure files
\usepackage{dcolumn}% Align table columns on decimal point
\usepackage{bm}% bold math
\usepackage{siunitx}
\usepackage[colorlinks=true,citecolor=blue,linkcolor=magenta]{hyperref}
\usepackage{amsmath}
\usepackage{enumitem}
\setlist{nosep} 

\hypersetup{
	pdfauthor = {Natascha Hedrich},
	colorlinks = true, linkcolor = blue, urlcolor=blue, bookmarksnumbered =  true}

\newcommand{\ket}[1]{\left|#1\right>}

\newcommand{\CrO}{Cr$_2$O$_3$}
\newcommand{\myJ}{\mathcal{J}}
\newcommand{\myK}{\mathcal{K}}
\DeclareMathOperator{\sech}{sech}
\DeclareMathOperator{\arcsinh}{arcsinh}

\begin{document}

\title{Nanoscale mechanics of antiferromagnetic domain walls}
\author{Natascha Hedrich}
\affiliation{Department of Physics, University of Basel, Klingelbergstrasse 82, Basel CH-4056, Switzerland}
\author{Kai Wagner}
\affiliation{Department of Physics, University of Basel, Klingelbergstrasse 82, Basel CH-4056, Switzerland}
\author{Oleksandr V. Pylypovskyi}
\affiliation{Helmholtz-Zentrum Dresden-Rossendorf e.V., Institute of Ion Beam Physics and Materials Research, 01328 Dresden, Germany}
\author{Brendan J. Shields}
\affiliation{Department of Physics, University of Basel, Klingelbergstrasse 82, Basel CH-4056, Switzerland}
\author{Tobias Kosub}
\affiliation{Helmholtz-Zentrum Dresden-Rossendorf e.V., Institute of Ion Beam Physics and Materials Research, 01328 Dresden, Germany}
\author{Denis D. Sheka}
\affiliation{Taras Shevchenko National University of Kyiv, 01601 Kyiv, Ukraine}
\author{Denys Makarov}
\affiliation{Helmholtz-Zentrum Dresden-Rossendorf e.V., Institute of Ion Beam Physics and Materials Research, 01328 Dresden, Germany}
\author{Patrick Maletinsky}
\email{patrick.maletinsky@unibas.ch}
\affiliation{Department of Physics, University of Basel, Klingelbergstrasse 82, Basel CH-4056, Switzerland}

\date{\today}

\begin{abstract}
Antiferromagnets offer remarkable promise for future spintronics devices, where antiferromagnetic order is exploited to encode information\,\cite{Jungwirth2018a,Jungwirth2016a,Baltz2018a}. 
The control and understanding of antiferromagnetic domain walls (DWs) - the interfaces between domains with differing order parameter orientations - is a key ingredient for advancing such antiferromagnetic spintronics technologies. 
However, studies of the intrinsic mechanics of individual antiferromagnetic DWs remain elusive since they require sufficiently pure materials and suitable experimental approaches to address DWs on the nanoscale. 
Here we nucleate isolated, $180^\circ$ DWs in a single-crystal of \CrO{}, a prototypical collinear magnetoelectric antiferromagnet, and study their interaction with topographic features fabricated on the sample. We demonstrate DW manipulation through the resulting, engineered energy landscape  and show that the observed interaction is governed by the DW's elastic properties.
Our results advance the understanding of DW mechanics in antiferromagnets and suggest a novel, topographically defined memory architecture based on antiferromagnetic DWs.
\end{abstract}

\maketitle

In the few years since its inception\,\cite{MacDonald2011a}, the field of antiferromagnetic spintronics\,\cite{Jungwirth2018a,Jungwirth2016a,Baltz2018a} has seen significant progress, culminating in several demonstrations of antiferromagnet-based memory devices\,\cite{Marti2014a, Wadley2016a, Kosub2017a}. 
While the focus of these advances has been on directly controlling and reading the bulk N\'eel vector of antiferromagnets\,\cite{Song2018a} and their domains\,\cite{Fiebig1995a}, the study and direct control of individual antiferromagnetic DWs 
has received much less attention 
thus far. 
Domain walls, however, are of particular relevance to the field, as they carry essential information on the magnetic microstructure of a material\,\cite{Hubert1998a,Weber2003a}, can have fundamentally different properties from the interior of domains\,\cite{Kummamuru2008a,Jaramillo2007a}, 
and delimit logical bits in magnetic memory devices\,\cite{Parkin2008a}. 
Furthermore, and akin to ferromagnet-based DW logic\,\cite{Allwood2005a,Luo2020a}, the understanding and control of antiferromagnetic DWs could inform novel approaches to antiferromagnetic spintronics architectures.

In this work, we realise a key step towards harnessing antiferromagnetic DWs for spintronics applications and thereby gain valuable insights into DW physics in antiferromagnets. 
Specifically, we realise an instance of antiferromagnetic DWs whose morphologies are governed by DW elasticity and sample geometry 
-- a key result we obtain by direct, real space imaging of antiferromagnetic DW trajectories with nanoscale resolution and over tens of micron lengthscales.

We demonstrate these results on the case of \CrO, a uniaxial, magnetoelectric antiferromagnet ordering at room temperature ($T_\textrm{N\'eel} =$ \SI{307}{K})\,\cite{Brown1969a}.
\CrO{}'s magnetoelectric properties allow for a direct, local control of the N\'eel vector $\bm{L}$, through the combined application of electric and magnetic fields\,\cite{Brown1969a,Wadley2016a}. 
Additionally, symmetry breaking at the surface of \CrO{} leads to a roughness-insensitive, uncompensated surface magnetic moment, that is directly linked to the underlying bulk N\'eel vector\,\cite{He2010a,Belashchenko2010a}, which can thereby be directly read out\,\cite{Kosub2015a}. 
These combined properties render \CrO{} particularly interesting for antiferromagnetic spintronics\,\cite{Kosub2017a,Jungwirth2016a}.

To address DW physics in \CrO{}, we employed nanoscale magnetic imaging using a single nitrogen vacancy (NV) electron spin in diamond as a scanning probe magnetometer\,\cite{Balasubramanian2008a, Rondin2014a}.
NV magnetometry is one of few nanoscale imaging methods for antiferromagnets\,\cite{Weber2003a,Bode2006a,Gross2017a,Appel2019a,Cheong2020a}, where it exploits stray magnetic fields resulting from uncompensated magnetic moments to address antiferromagnetic order. Such moments can generally result on surfaces\,\cite{Belashchenko2010a} or from spatial variations of the N\'eel vector\,\cite{Andreev1980a,Bode2006a,Tveten2016a} 
and are thereby particularly suitable for studying antiferromagnetic DWs.
Here, we exploit \CrO{}'s surface magnetic moments\,\cite{He2010a} for imaging.

\begin{figure*}[t]
         \includegraphics[width=17.8cm]{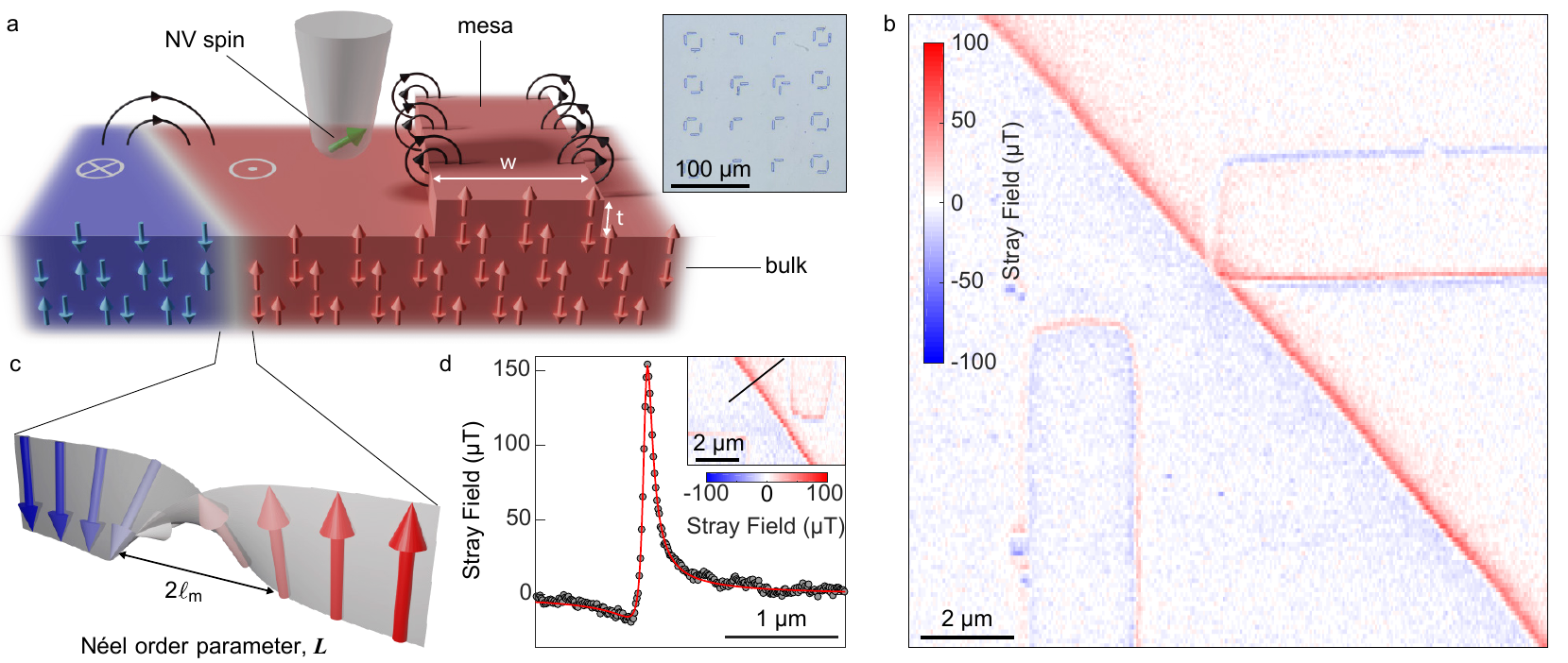}
	\caption{\label{fig:Sample}
	 \textbf{Sample structure and domain wall imaging on single crystal antiferromagnetic \CrO{}.} 
	 \textbf{a}~Schematic of the (0001)-oriented \CrO{} single crystal sample, showing the spin structure of two  sections with oppositely oriented N\'eel vector $\bm{L}$ and the associated surface magnetisation. Stray magnetic fields occur either on domain walls (DWs) or on topographic features and are measured and imaged using scanning single spin magnetometry (see text). The inset shows a micrograph of the mesas (of thickness $t\approx166~$nm) fabricated on the sample surface.
	\textbf{b}~Representative stray-field image obtained on the surface of the sample on a section containing two mesas and a DW nucleated by magnetoelectric cooling (see text). 
	\textbf{c}~Schematics of the N\'eel vector evolution across the DW of width $2\ell_m$. 
	\textbf{d}~Line-cut of the magnetic field measured across the antiferromagnetic DW. From a fit to the data (red), we determine an upper bound for the magnetic length $\ell_m\lesssim 32~$nm.
	}
\end{figure*}

We performed our experiments on a $(0001)-$oriented \CrO{} single-crystal, \SI{1}{\mm}-thick, with millimetre-scale lateral dimensions. To obtain position markers and a measurable magnetic stray field from the sample, even from uniformly ordered domains,  we pattern a grid of micron-scale mesas (Fig.\,\ref{fig:Sample}a, inset) with mean thickness $\bar{t}=\SI{166\pm4}{\nm}$ and width $\bar{w}=\SI{2.4\pm0.3}{\um}$ on the sample surface using standard lithography (methods).  To induce magnetic domains in the \CrO{} sample, we employ magnetoelectric field cooling\,\cite{Brown1969a,Borisov2005a} across $T_\textrm{N\'eel}$. 
Specifically, we apply collinear electric and magnetic bias fields along the surface normal with $B_{\rm bias}=550~$mT and $E_{\rm bias}=\pm0.75~$MV/m, where we used a split-gate capacitor to invert $E_{\rm bias}$ between two halves of the sample. 
We find this method of nucleation to be repeatable, reversible and necessary to observe DWs in the otherwise mono-domain sample (SI).

Figure\,\ref{fig:Sample}b shows a representative NV magnetometry image of a section of the sample obtained at room temperature and in a weak bias magnetic field ($B_{\rm{NV}}=1.6~$mT) applied along the NV axis to achieve quantitative imaging\,\cite{Rondin2014a}. 
The data show stray magnetic fields emerging from a nucleated DW and from two mesas located on adjacent antiferromagnetic domains.  
From an analytical fit to the stray field across the mesa edges, we extract mean surface magnetizations $\sigma_m=\pm$\SI{2.1 \pm 0.3}{\mu_B/ nm^2} 
(where the two signs apply to the different mesas), 
consistent with previous measurements\,\cite{Brown2002a, Appel2019a} and theoretical expectations\,\cite{Shi2009a}. 
We confirm that these data are connected to the bulk antiferromagnetic order by performing temperature-dependant measurements, $\sigma_m(T)$, where we observe that $\sigma_m$ vanishes near $T_\textrm{N\'eel}$\,\cite{Brown1969a}. 
Details on the fitting routines and data for $\sigma_m(T)$ are given in the SI.

The DW we observe constitutes an interface between regions of oppositely aligned N\'eel vector, $\bm{L}$, through which $\bm{L}$ rotates by $180^\circ$ over a characteristic lengthscale $2 \ell_m=2\sqrt{\mathcal{A}/\myK}$ (Fig.\,\ref{fig:Sample}c), where $\mathcal{A}$ and $\myK$ are the exchange stiffness and magnetic anisotropy\,\cite{Hubert1998a}. 
Using our measured value of $\sigma_m$ as an input parameter, we perform an analytical fit to the magnetic stray field measured across the DW (Fig.\,\ref{fig:Sample}d, inset) to determine a room temperature upper bound of $\ell_m\lesssim32~$nm, consistent with theoretical estimates (see SI).
Strikingly, we find the DWs away from the mesas to be largely smooth and straight over length scales of tens of microns (see Fig.\,\ref{fig:Sample}b and SI for further representative DW images), and do not observe a correlation between DW orientation and crystallographic directions of the sample.

\begin{figure}[t]
	\includegraphics[width=8.6cm]{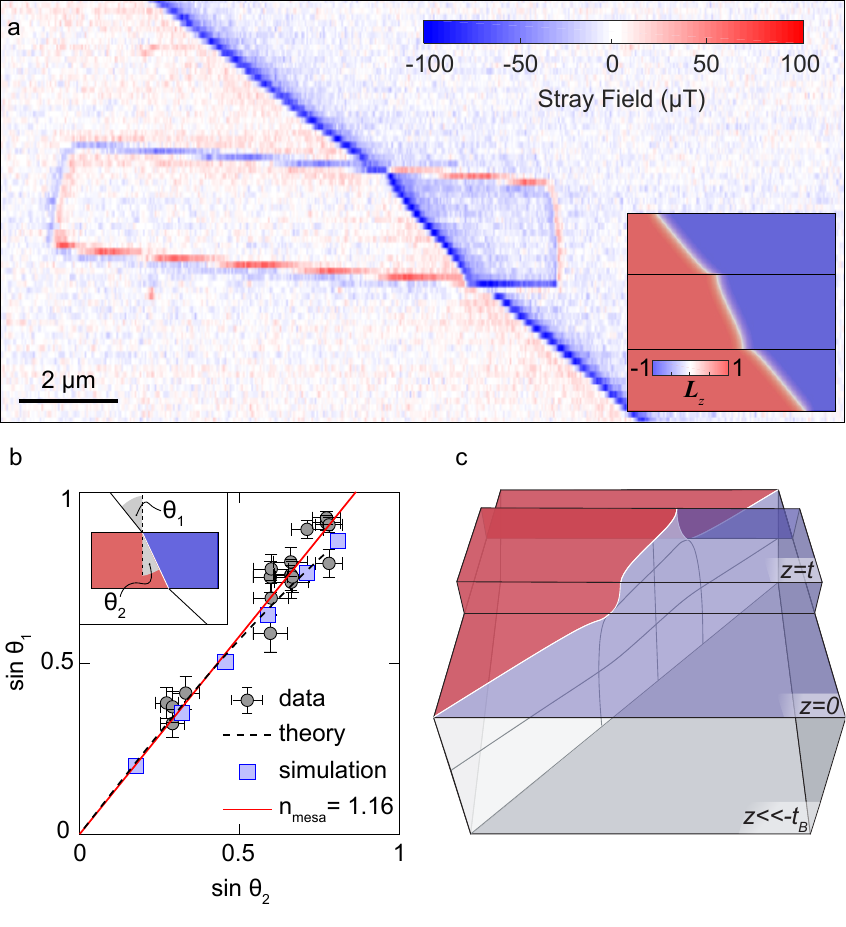}
	\caption{\label{fig:Kink} \textbf{Mechanics of an antiferromagnetic domain wall.} 
	\textbf{a}~Stray magnetic field image of an antiferromagnetic domain wall (DW) crossing a mesa. The mesa deflects the DW from its otherwise straight path and leads to a further DW distortion within the mesa.  
	Inset: The DW trajectory as found by a numerical simulation (see text) shows excellent agreement with data.
	\textbf{b}~Sines of the incidence and transmission angles ($\theta_{\rm{1}}$ and $\theta_{\rm{2}}$ respectively, as defined in the inset) for a DW incident on a mesa edge, as determined from $17$ DW images (grey circles). The apparent linear relationship (red line) is reminiscent of Snell's law in geometrical optics and suggests DW energy minimisation as the origin or the observed deflection. Numerical simulations (blue squares) and analytic calculations (dashed line) confirm the experimental findings for mesa aspect ratios $\approx \bar{t}/\bar{w}$. 
	\textbf{c}~Full, three-dimensional representation of the simulated DW surface crossing a mesa. 
	Below the mesa, the DW twists towards the planar DW surface in the bulk over a characteristic lengthscale $t_B$. 
	}
\end{figure}

When a DW crosses a mesa, however, we observe considerable deviations from such straight DW paths,  which bear similarity to the refraction of a light beam as described by Snell's law in geometrical optics (Fig.\,\ref{fig:Kink}a). 
Indeed, 
the crossing of the DW through the mesa incurs an energy cost, directly proportional to the increase in DW surface area resulting from the non-zero mesa height $t$.
The DW then assumes a path that minimizes its surface area (and thereby its surface energy), taking into account the local change in topography.  
To further support this picture, we mapped $17$ instances of such refraction-like behavior for a wide range of DW incidence angles, $\theta_{\rm 1}\in\{\sim20^\circ...\sim70^\circ\}$, as summarized in Fig.\,\ref{fig:Kink}b. 
We determine $\theta_{\rm 1}$ by the DW direction off the mesa and define the outgoing angle $\theta_{\rm 2}$ from the DW direction at the center of the mesa (Fig.\,\ref{fig:Kink}b, inset). 
Similar to Snell's law, we find a linear behavior 
$\sin\theta_{\rm 1}/\sin\theta_{\rm 2}=1.16\pm0.04$.

To obtain further insight into the observed DW mechanics, we perform spin lattice simulations\,\cite{SLaSi,Pylypovskyi13f} that take into account nearest-neighbour antiferromagnetic exchange interactions, single-site anisotropy and the sample geometry (see Methods). 
We then obtain the equilibrium DW configuration through energy minimization of the spin lattice. 
The simulated DW profile on the sample surface (Fig\,\ref{fig:Kink}a, inset) 
shows excellent agreement with the experimental data. 
Extracting $\sin\theta_{\rm 2}$ from our simulations for varying values of $\sin\theta_{\rm 1}$ confirms the experimentally observed linear relationship (Fig.\,\ref{fig:Kink}b).

Our numerical results inspired an analytic ansatz for the DW profile, where we use a variational procedure to relate key parameters of the DW morphology to the mesa geometry (see SI). This analysis yields an analytic expression for $n_\text{mesa} := \sin \theta_1 / \sin \theta_2$ (Fig.~\ref{fig:Kink}b, dashed line), where for small angles $\theta_1 \ll 1$ 
we find 
$n_\text{mesa} = 1 + 3.1(t/w)$ and additional terms $\mathcal{O}(\theta_1 ^2)$. Although Snell's law offers a useful analog to the observed phenomena, this result 
also highlights distinctions between the two. In particular, while the former arises from the principle of least action alone, the DW trajectory is additionally determined by the DW position in the bulk of the sample, far from the surface, which manifests in the existence of higher-order contributions of $\theta_1$ to $n_{\rm mesa}$. 

Strikingly, the simulated DW trajectories also reproduce the marked bending of the DW towards the mesa edge normal, resulting in an "S-shaped" distortion from the otherwise straight DW profile (Fig.\,\ref{fig:Kink}a). 
This distortion arises from the minimization of the exchange interaction by normal incidence of DWs to surfaces (in this case, the mesa edge)\,\cite{Hubert1998a}. This, together with the overall DW energy minimization, fully explains the observed DW trajectory on the mesa. 
Our simulation also yields the full, three dimensional morphology of the DW crossing the mesa (Fig.\,\ref{fig:Kink}c), and shows how the mesa-induced distortion of the DW transitions towards the planar, bulk DW shape over a characteristic lengthscale $t_B$. Through our analytic analysis, we find $t_B \approx 0.34 w$, which yields $t_B$ = \SI{0.82}{\um}.

\begin{figure*}[t]
\includegraphics[width=17.8cm]{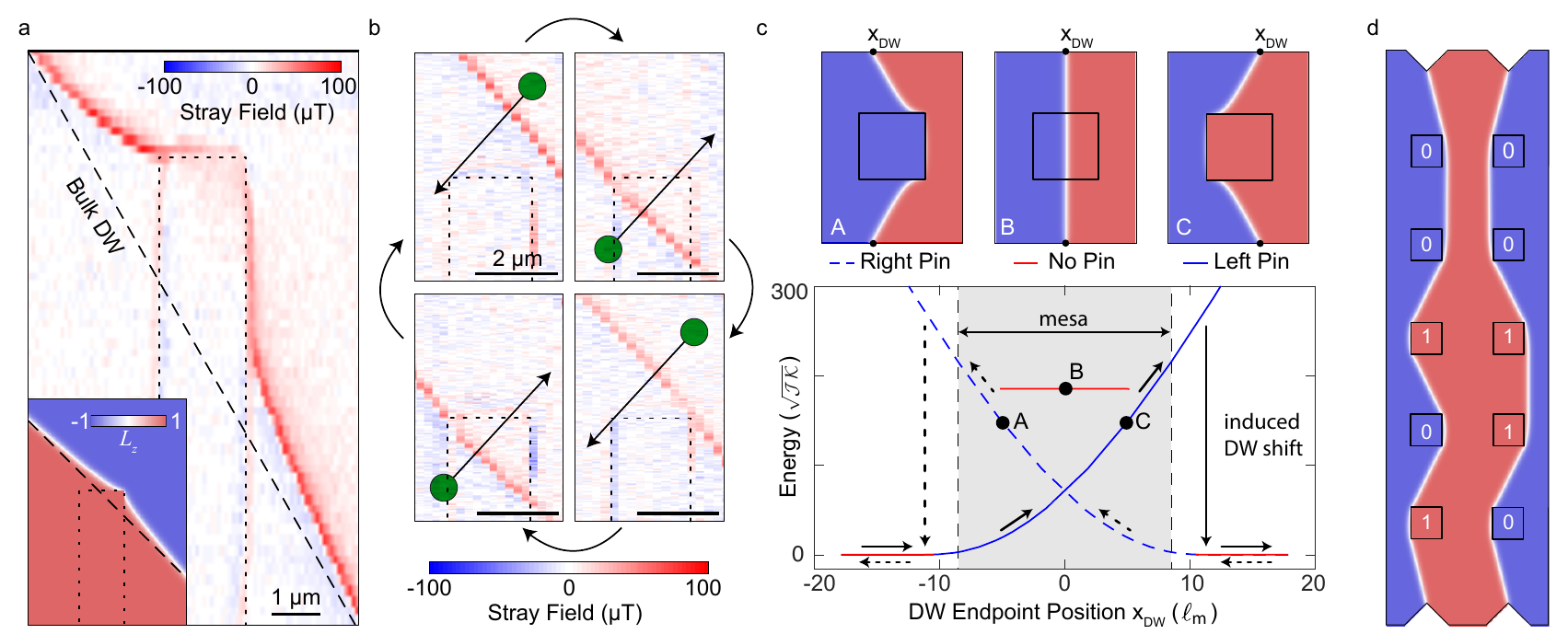}
	\caption{\label{fig:Memory} \textbf{Engineered pinning and controlled manipulation of antiferromagnetic domain walls} 
	\textbf{a}~Magnetic stray field map of a domain wall (DW) pinned to a mesa corner. The inset shows a corresponding simulation. 
	\textbf{b}~Laser dragging of a DW across a second mesa corner (unique from that in \textbf{a}). Green dots and black arrows show the path along which the DW is dragged. The mesa position is highlighted by the black, dashed line. 
	\textbf{c}~Top: Simulation snapshots of the three possible, (meta-)stable DW states in the vicinity of the mesa: Pinned to either side of the mesa or running straight across the mesa. Bottom: Energy of the three states (color code defined above) as function of DW position, as parameterised by $x_\textrm{DW}$, the fixed DW position away from the mesa (see top panels). An externally applied stimulus can cause the DW to relax to an energetically favourable state (vertical arrow). 
	\textbf{d}~Proposal for a DW-based antiferromagnetic memory. Bit locations are defined by the mesa structures (black squares) and information is encoded by the direction of $\bm{L}$ on the mesa surface.
	}
\end{figure*}

The energy-penalty for traversing a mesa also leads to 
DW pinning phenomena at mesa edges.  
Specifically, we observe instances where the bulk DW position would intersect a mesa close to a corner, but is expelled from the mesa to minimise DW energy (Fig.\,\ref{fig:Memory}a).
This behavior is well reproduced in simulations, where we force the bulk DW to lie close to a mesa corner (Fig.\,\ref{fig:Memory}a, inset). 
In such a case, the mesa presents a large DW energy barrier, and so the path, which minimises the overall energy, follows the mesa edge.  
The energetically favourable DW path (``refraction'' or ``pinning'') is therefore dependent on the mesa geometry 
and the location of the DW with respect to the mesa.

Importantly, we are able to de-pin the DW from a mesa corner 
and to place it on the mesa by using a focused laser spot to drag the DW (Fig.\,\ref{fig:Memory}b).
Such laser-dragging, previously demonstrated in the case of thin-film ferromagnets\,\cite{Tetienne2014a}, is based on laser-induced heating, which locally reduces the DW energy and therefore forms a movable potential well for the DW. 
We find that such local DW manipulation is facilitated by increasing the sample temperature, and therefore heat the sample to  
$304.5~$K.   
We then scan the laser at elevated powers, perpendicular to the DW and in absence of the scanning probe. 
As shown in Fig.\,\ref{fig:Memory}b, we are able to reproducibly move the DW from lying on the mesa to a nearby location off the mesa and back, and thereby demonstrate the ability to achieve reliable and reversible DW manipulation.

We investigate this pinning and switching behavior more closely through simulations. In particular, we consider the energy of a DW whose end points we pin at fixed positions $x_\textrm{DW}$ relative to a square mesa (Fig.\,\ref{fig:Memory}c). By varying $x_\textrm{DW}$, we observe three distinct equilibrium DW configurations: a straight, undisturbed DW, or the DW pinned to either side of the mesa. The energy and snapshots of these configurations are shown as a function of $x_\textrm{DW}$ in Fig.\,\ref{fig:Memory}c. In the case of a straight DW, the DW energy increases abruptly upon crossing the mesa, where the energy-step can be tuned by the mesa height. For the pinned DWs, the DW area, and hence the DW energy, increases gradually with increasing DW deflection. In particular, this observed pinning behavior shows that the DW mimics an elastic surface, which can be deformed. The mechanical surface tension of the DW (energy per unit area) can be determined by the material properties and determines the energy excess resulting from the interaction between the DW and mesa (see SI). For sufficiently strong deflections, the energy of the pinned state exceeds that of a straight DW, leading to a metastable configuration. When this happens, applying a stimulus (e.g. magnetoelectric pressure or local heating) can cause a sudden change in the DW state (vertical arrows in Fig.\,\ref{fig:Memory}c). This switching process is hysteretic and controllable by the mesa-height and the strength of the stimulus.

These combined results suggest a novel architecture for a scalable, DW-based antiferromagnetic memory as outlined in Fig.\,\ref{fig:Memory}d. 
Specifically, we propose to employ nanoscale mesas as engineered DW pinning centers, where binary information is encoded by the direction of $\bm{L}$ on a given mesa. 
By fabricating sufficiently thick mesas, one can energetically exclude the case of an unpinned DW, thereby creating a bistable system where the DW is forced to pin to one edge of the mesa or the other. The size of the resulting antiferromagnetic memory bits is then limited by $\ell_m$ only, opening the route to bits of nanoscale dimension -- a significant improvement on currently demonstrated architectures for antiferromagnetic memories\,\cite{Wadley2016a,Kosub2017a}.
We demonstrated the possibility to switch and read such bits by laser dragging and magnetometry, but integrated, all-electrical approaches could be readily envisaged. 
In particular, electrical gates on the mesas could be used to apply magnetoelectric pressure\,\cite{He2010a,Ashida2015a} for switching, and to exploit the anomalous Hall effect\,\cite{Kosub2015a} for all-electrical readout.

In conclusion, we demonstrated the deterministic generation and control of pristine DWs in a single-crystal antiferromagnet, and observed DW physics determined solely by sample geometry and the DW surface energy, while defect-related DW pinning mechanisms\,\cite{Lemerle1998a} appear negligible.
These combined achievements, together with the versatile toolset of quantum sensing\,\cite{Degen2017a}, 
offer attractive avenues to exploring largely uncharted areas of DW physics, such as DW creep\,\cite{Chauve2000a,Ferre2013a} or DW magnons 
in antiferromagnets\,\cite{Flebus2018a}.  
Based on the generality and robustness of our modeling (see Methods), we conclude that our results should be extendable to other achiral, uniaxial antiferromagnets and open multiple avenues for future research of fundamental and applied nature, be it in the form of the proposed antiferromagnetic memory devices, or ultimately for the realisation of DW logic\,\cite{Allwood2005a}. Through simulations, we expect the same behavior seen here in bulk, to manifest in high-quality, thin-film samples -- a key materials frontier that remains to be addressed in future work. 

\section{Methods}
\subsection{Sample Preparation}
The \CrO{} used in this study is a commercially available, single crystal from MaTeK with a (0001) surface orientation. The originally \SI{5x5x1}{\mm} crystal was broken into two halves along a diagonal (see SI). The sample was prepared by removing magnetic contamination (presumably resulting from the polishing process) with a \SI{100}{\s} ArCl$_2$ plasma etch (ICP-RIE, Sentech) in \SI{2}{\s} steps. One side of the crystal was then spin coated with an HSQ layer (FOx, Dow Corning) and subsequently developed using electron-beam lithography to create \SI{10x2}{\um} mesa masks. These mask patterns were transferred into the sample with a \SI{100}{\s} ArCl$_2$ plasma etch. The masks were then removed using HF. This process results in the \SI{166}{\nm}-tall structures seen in the inset of Fig.~\ref{fig:Sample}a. For the measurements, the sample was mounted on a small Peltier element, placed on top of an open-loop, piezoelectric scanner (Attocube ANSxyz100), allowing us to heat the sample up to $\approx$\SI{340}{K}.

\subsection{NV Magnetometry}
The NV center is a point defect in diamond, whose S=1 electronic ground state spin can be initialized and read-out through optical excitation at \SI{532}{\nm}. Specifically, we use state-dependent fluorescence to identify the Zeeman splitting between the $\ket{\pm1}$ spin levels using optically detected magnetic resonance (ODMR).
All measurements in this study were performed using scanning, all-diamond parabolic pillars~\cite{Hedrich2020a} housing a single NV center and integrated into a custom confocal imaging setup equipped with a CW \SI{532}{\nm} laser~\cite{Grinolds2013}. The measurements were performed with $<$\SI{10}{\uW} of continuous-wave optical excitation, a factor of two smaller than typical saturation powers for NVs in these parabolic scanning pillars~\cite{Hedrich2020a}. The microwave (MW) needed to manipulate the NV is provided by a \SI{30}{\um} gold loop antenna with a typical effective driving strength of \SI{0.25}{G} at the NV. These low excitation powers (both MW and laser) ensure that we do not disturb the DW, which was confirmed by repeating scans and observing the DW. A small bias magnetic field ($<$\SI{60}{G}) was applied along the NV axis using a permanent magnet to allow for a sign-sensitive measurement of the stray magnetic fields. 

Both 2D magnetic field images and linescans were performed using a feedback technique to lock a microwave driving frequency to the instantaneous NV spin transition frequency, as described in~\cite{Schoenfeld2011}. We employ single-pixel integration times ranging from \SI{0.3}{\s} for full-field images to \SI{5}{\s} for individual line scans with a noise floor of $\approx$~\SI{3.3}{\micro\tesla}$/\sqrt{\textrm{Hz}}$.

\subsection{Domain Wall Nucleation}
DWs are nucleated in the otherwise mono-domain single crystal \CrO{} using magnetoelectric cooling through the N\'eel temperature. A uniform magnetic field is achieved by placing two \SI{5x5}{\cm} permanent magnets adjacent to each other and in close proximity to the sample. The result is a nearly homogeneous magnetic field of $B\approx550~$mT along the center normal, as measured by a Hall probe (AS NTM, FM302 Teslameter, Projekt Elektronik). In addition, we apply electric fields across the sample using a split-gate capacitor consisting of two quartz plates with \SI{100}{\nm} Au evaporated onto the surface. 
The \CrO{} sample is then centered onto the capacitor gap and sandwiched between the top and bottom gate together with thin mica sheets to prevent electroplating of Au onto the crystal surface. A schematic of this setup may be found in the SI. The whole device is then heated to far above the N\'eel temperature and allowed to cool to room temperature while simultaneously applying $\pm$\SI{750}{V} between the electrodes leading to an electric field of $E\approx$ \SI{0.75}{MV/m} across the crystal. The resulting $|\bm{E}\times \bm{B}| = 0.41\times10^6$ VT/m is sufficient to force the \CrO{} sample into a two-domain state. The crystal can again be made mono-domain by repeating this procedure with a uniform capacitor rather than the split-gate. This process has been repeated twice to show the reproducibility, where each realization resulted in a different domain configuration (see SI). 

\subsection{Domain Wall Dragging}
The repeated movement of the DW is demonstrated via local heating with a laser - a process we describe as "laser dragging". For this, we remove the NV scanning probe and focus the laser onto the sample surface with a beam diameter of $\approx$\SI{420}{\nm}. We scan the sample at a speed of roughly \SI{80}{\nm/s}, perpendicular to the DW, over distances exceeding \SI{10}{\um} before reducing the laser power back to below \SI{10}{\uW}. With this method, the minimum laser power at which we have observed DW motion is \SI{135}{\uW}. We then replace the NV scanning probe and image the new DW position. These experiments were performed at a sample temperature of $304~$K, achieved through heating with the Peltier element. This domain wall dragging can also be verified through direct measurements with the NV (see SI).

\subsection{Fitting to Mesa Stray Fields}
The mesa structures etched into the surface of \CrO{} play a critical role in our study as they act as sources of stray fields for characterizing the surface magnetization ($\sigma_m$) and NV-sample spacing ($d_{NV}$) as well as providing reference markers. For the former, we consider 29 linecut sections, each taken over a mesa, and fit the stray field at the mesa edges by modifying a well-studied model~\cite{Tetienne2015}, which describes the stray field as arising from line currents along the top and bottom edges of the mesa (see SI). We obtain estimates of the NV angles ($\theta_{NV}$ and $\phi_{NV}$), $d_{NV}$, $\sigma_m$ and the mesa edge positions, as well as their variances, through the Metropolis Hastings (MH) algorithm (see SI). In particular, by multiplying the likelihood distributions of all 29 datasets, performed at various temperatures, we obtain reasonable estimates of the NV angles, $\theta_{NV} = 60.7\pm2.9$ and $\phi_{NV} = 260.6\pm 0.8$ degrees. We furthermore extract a mean $d_{NV}=51.4\pm19.2~$nm. By considering only the six measurements taken at room temperature, we determine the value for the surface moment density, $\sigma_m = 2.1 \pm 0.3 $ $\mu_B$/nm$^2$, where the error corresponds to the standard deviation of the measurements. 

\subsection{Fitting to Domain Wall Stray Fields}
To describe the stray field of a domain wall in \CrO, we begin with the typical description of the evolution of the magnetic moments of the two sublattices in this collinear antiferromagnet \cite{Belashchenko2016a,Mitsumata2011}. 
For this, we assume a Bloch wall of the typical form:
\begin{align}
L_x &= 0,\\
L_y &= \sech(x/\ell_m),\\
L_z &= \tanh(x/\ell_m),
\label{eq:DWprofile}
\end{align}
where $\ell_m$ is the magnetic length, as given in the main text. Thus, the domain wall profile (as shown in Fig.~\ref{fig:Sample}c) is determined by $\ell_m$, allowing us to use this parameter to characterize the domain wall width. This form of the domain wall profile is then considered in the derivation of the stray field along the NV axis, as described in the SI. 

We again use the MH algorithm to evaluate our model of the stray field with the stray field data. To do so, the NV-sample spacing, angles and surface magnetization previously extracted from the mesa fits are used as prior information in the fit. In particular, we consider the NV-sample spacing on a case-by-case basis as each DW linescan is taken concurrently with a mesa linescan. The upper bound for $\ell_m$ stated in the main text is then obtained from the extracted likelihood distributions at room temperature. We examine the extrema of the distributions, selecting the 98$^{th}$ percentile of the cumulative distribution function (CDF) as the upper limit on $\ell_m$. This implies that, at room temperature, our data excludes an $\ell_m$ larger than \SI{32}{\nm}. 

We note that for completeness, the strayfield data has also been analyzed under the assumption of a N\'eel wall, resulting in a similar quality fit for slightly changed model parameters. However, the resulting domain wall width is consistently smaller for a N\'eel wall, verifying the validity of our statement on the upper limit of $\ell_m$, regardless of wall type.

\subsection{Simulation Details}
The spin-lattice simulations are performed in the in-house developed SLaSi package~\cite{SLaSi, Pylypovskyi13f}, rewritten in the CUDA framework, and based on a generic antiferromagnet consisting of a simple cubic lattice, described by the effective Hamiltonian:
\begin{equation}
\begin{aligned}
\mathcal{H} & = \dfrac{\myJ S^2}{2} \sum_{i,i'}  \bm{\mu}_{i} \cdot \bm{\mu}_{i'} - \dfrac{K S^2}{2} \sum_{i} (\mu_{i}^z)^2\\
&+ c_d \dfrac{\mu_0g^2\mu_\textsc{b}^2S^2}{8\pi} \sum_{i\neq j} \left[ \dfrac{\bm{\mu}_i\cdot \bm{\mu}_j}{r_{ij}^3} - 3 \dfrac{(\bm{\mu}_i\cdot \bm{r}_{ij})(\bm{\mu}_j\cdot \bm{r}_{ij})}{r_{ij}^5} \right].
\end{aligned}
\label{eq.ham}
\end{equation}
Here, $\myJ$  is the exchange integral, $K$  is the easy-axis ($\bm{e}_z$) anisotropy, $\bm{\mu}_i$ is the unit vector representing the direction of the magnetic moment at the $i$-th lattice site and $i'$ runs over the nearest neighbors of $i$, yielding the oppositely oriented magnetic sublattices. To represent a general antiferromagnet, all while approximating the properties of \CrO{}, we set $S=1$, $a = 0.277~$nm, $\myJ$= 2.34$\times 10^{-9}~$pJ~\cite{Shi2009a} and $K$= $2.6\times 10^{-10}~$pJ, which leads to $\ell_m = a\sqrt{\myJ/K} =$ \SI{0.83}{\nm}. These values allow for a reasonable scale of the sample for spin-lattice simulations and properly reproduce the effects observed in experiments. The last term in Hamiltonian (\ref{eq.ham}) represents the dipolar interaction, which we control by the parameter $c_d\in\{0,1\}$. We find that dipolar interactions do not change the results of our simulations qualitatively nor quantitatively, and therefore favor $c_d=0$ in the following, consistent with other studies.

With this, we solve the set of Landau-Lifshitz-Gilbert equations
\begin{equation}
\dfrac{\mathrm{d}\bm{\mu}_i}{\mathrm{d}t} = \dfrac{1}{\hbar S} \bm{\mu}_i \times \dfrac{\partial \mathcal{H}}{\partial \bm{\mu}_i} + \alpha_\textsc{g} \bm{\mu}_i \times \dfrac{\mathrm{d}\bm{\mu}_i}{\mathrm{d}t},
\end{equation}
where $\alpha_\textsc{g} = 0.5$ is the Gilbert damping, using the Runge-Kutta-Fehlberg scheme of order 4-5 with a fixed time step to find the equilibrium magnetic state when $\max|\mathrm{d}\bm{\mu}_i/\mathrm{d}t|\to 0$. We simulate parallelepiped-shaped samples with the mesa faces coinciding with lattice planes. This is done without loss of generality, as simulations with arbitrarily oriented mesas show no significant variations. 

To simulate a given bulk domain wall position, in particular for the study shown in Fig.~\ref{fig:Memory}c, we fix the equilibrium domain wall by notches at the sample boundaries. The initial state is defined as either a straight DW, which can cross the mesa, or one which is pinned at and bent around the mesa edges. The magnetization is then relaxed and its energy is compared to that of an unperturbed domain wall far from the mesa. The excess energy of the initial state can cause the DW to switch from a high energy (strongly extended) to low energy state, thereby imitating an induced switch due to an external stimulus. We note that the present model contains no bias to select a particular DW type (Bloch, N\'eel, or other). The resulting equilibrium DW type in the simulations is thus determined by the initial magnetic state we choose before numerical relaxation and may also vary along the DW. We varied details of the initial conditions and observed no influence of the DW type on the relaxed DW trajectory.

Finally, to further investigate the robustness of our model, we performed simulations where we lower the structural symmetry of the lattice by shifting the two magnetic sublattices we consider by half a lattice constant with respect to each other along the main axis of the cubic lattice. Here, we also observe quantitatively similar results as for the original, simple cubic lattice, indicating that our model is indeed robust against variations in model parameters. Thus, though the considered spin-lattice is not a perfect representation of the \CrO{} spin-lattice, these simplifications appear justified as our minimal model already captures all features of DW mechanics observed in our experiments. Furthermore, the generality of this model means it should be applicable to any achiral, uniaxial antiferromagnet.

\section{Acknowledgements}
We thank O.~Gomonay and S.A.~D\'{i}az for fruitful discussions and M.~Fiebig and M.~Giraldo for optical characterisation of our \CrO{} samples at an early stage of the experiment. 
We would further like to thank M.~Kasperczyk and P.~Amrein for their help with efficient implementations of the Metropolis-Hastings algorithm, A.~K\'{a}kay at the HZDR for providing us with computation time for micromagnetics, and D. Broadway and L. Thiel for valuable input on figures. We would also like to thank A.V.~Tomilo at the TSNUK for his help with the spin-lattice simulations as well as his very helpful insight.

We gratefully acknowledge financial support through the NCCR QSIT, a competence centre funded by the Swiss NSF, through the Swiss Nanoscience Institute, by the EU FET-OPEN Flagship Project ASTERIQS (grant $820394$), SNF Project Grant $188521$, German Research Foundation (projects MA $5144/22-1$ and MA $5144/24-1$) and Taras Shevchenko National University of Kyiv (Project No. $19$BF$052-01$). 

\bibliographystyle{apsrev4-1}
\bibliography{BibSCCr2O3}

\clearpage

\section*{Supplementary Information}

\section{Position Calibration and Error Analysis} \label{sec:error}
\begin{figure*}[th]
		\centering
		\includegraphics[width=1\textwidth]{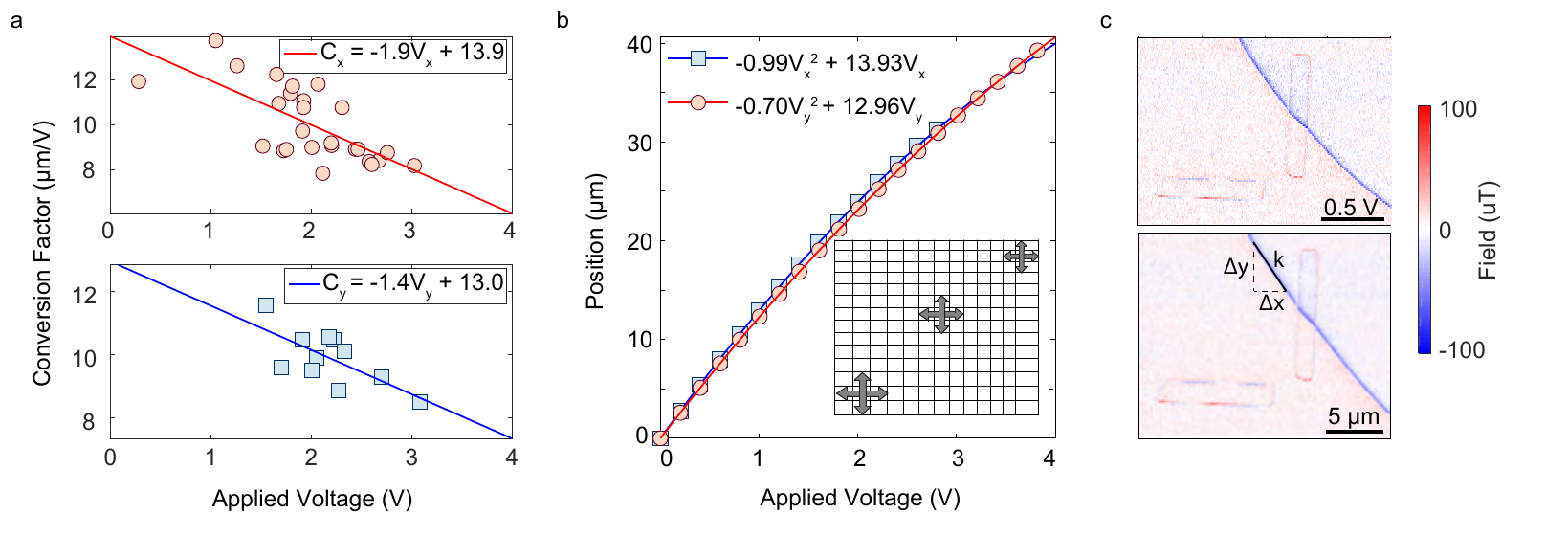}
		\caption{\textbf{Calibration of the piezo position} \textbf{a} The conversion factor from applied voltage (V) to displacement (\SI{}{\um}) as a function of the applied voltage for the x (top) and y (bottom) axes. \textbf{b} Plot of the position (as determined by integrating the conversion factor in \textbf{a}) as a function of the voltage applied in the x direction (blue squares) and y direction (red circles). The fit parameters are given in the top right inset. The bottom right inset shows the resulting deformation of equally spaced lines in a 2D plot, with dark gray arrows shown for emphasis. \textbf{c} Stray field image of two mesas together with the domain wall plotted with the position in applied voltage (top) and the converted position in \SI{}{\um} (bottom).}
	\label{fig:calib}
	\end{figure*}

	As we are using open-loop piezo scanners (Attocube ANSxyz100), we need to calibrate their physical displacement and determine the piezo non-linearity in order to achieve accurate fitting of the domain wall. To do so, we perform atomic force microscopy (AFM) measurements of our sample's topography on a commercial system (Bruker Dimension 3100), and compare various length scales as offered by patterned mesas on the sample surface to those measured by our system. This allows us to determine the conversion factor from applied voltage (V) to physical piezo displacement (\SI{}{\um}) for a wide range of piezo voltages as shown in Fig.~\ref{fig:calib}a. Errors on individual points here are below the marker size. We integrate the fitting functions given in the legends to convert from our system coordinates (in voltage) to real coordinates (in \SI{}{\um}) as shown in Fig.~\ref{fig:calib}b. This leads a non-linear conversion, which corrects for deformations in our measurements (Fig.~\ref{fig:calib}b, inset). We make this explicit by showing a 2D stray field image in Fig.~\ref{fig:calib}c, where the original image (with scale in applied voltage) is shown in the top panel and the adjusted scaling conversion in the bottom panel. The difference between the two images is most apparent when one compares the visible curvature of the domain wall in both figures.  
	\\ From these conversion factors, we can also obtain an estimate of the uncertainty in our displacement calibration. We compare the mesa dimensions measured in our setup to those measured with the Bruker Dimension 3100 AFM % as well as characterizing our piezo movement with an interferometer system (Attocube IDS), 
	and obtain a 10\% error. We therefore assume a 10\% uncertainty on quantities such as the $\Delta x$ and $\Delta y$ that factor into the error bars in Fig. 2b in the main text. This uncertainty will also appear in the mesa width used to determine the magnetization and the NV-to-sample separation. 
	\\Finally, let us further examine the error analysis for the domain wall (DW) deflection angles. Assuming  $\Delta x = \Delta y = 10\%$, we obtain $\delta k_{(1)}  = \sqrt{2}k_{(1)}\Delta x$, where $k~(k_1)$ is the slope of the domain wall in the bulk (on the mesa) (see Fig.~\ref{fig:calib}c), which can then be converted to an error on $\sin(\theta_{1(2)})$ through simple error propagation. This results in
	\begin{equation}
	\delta\sin(\theta_{1(2)}) = \frac{\sqrt{2}\cos(\tan^{-1}(k_{(1)}))k_{(1)}}{1+k_{(1)}^2}\Delta x,
	\end{equation}
	which is plotted as the error bars in Fig. 2b of the main text. We then use a linear regression, to obtain the final estimate of $n_{\textrm{mesa}}$. The remaining error analysis will be addressed in later sections. 

	\section{Domain Wall Nucleation and Morphology} \label{sec:nucleation}
\begin{figure*}[th]
		\centering
		\includegraphics[width=0.8\textwidth]{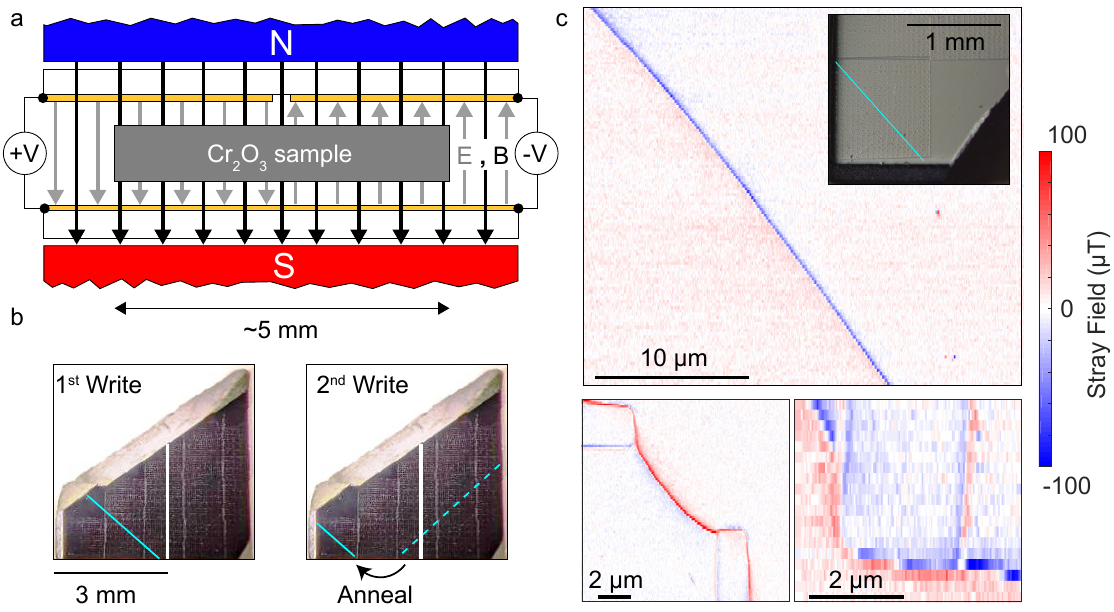}
		\caption{\textbf{Nucleation of domain walls} \textbf{a} A schematic representation of the setup used to nucleate domain walls in a single crystal of \CrO. The north and south poles of the permanent magnets are shown with the field direction given by the black arrows. The grey arrows show the direction of the electric field, generated by applying voltages $\pm$V between the top contacts and ground. \textbf{b} Optical images of the crystal with the approximate locations of the split-gate gap during nucleation shown with a thick, white line. The final domain wall position is given by the solid, cyan line. In the left image, the dashed cyan line shows the position of the domain wall before annealing. \textbf{c}. (top) Stray field image of the domain wall showing its general smoothness and straight orientation. The inset shows a microscope image of the domain wall position, shown by the solid, cyan line, with the area of the stray field scan taken at the bottom right end of the line.  (bottom) Stray field images showing (left) simultaneous pinning at two different mesa edges after the first poling and (right) a further example of DW pinning to two edges of a mesa after the second poling.}
	\label{fig:DW}
	\end{figure*}

	As received, the single crystal \CrO{} was in a mono-domain state, confirmed with high probability by measuring the same sign of the surface magnetization at number of mesas (see Section~\ref{sec:mesafit}) across the surface of the crystal. The DW is then nucleated as described in the methods and using the device shown in Fig.~\ref{fig:DW}a. We use the same method of sampling the magnetization across the sample to localize the DW. The nucleated DWs appear smooth and straight, as seen when imaging the domain wall over larger areas. The approximate orientation of the DW, as determined by NV magnetometry, is shown in Fig.~\ref{fig:DW}b with cyan lines for subsequent nucleation procedures, separated by an erasure of the domain wall through electromagnetic cooling in a homogeneous field. Note that these positions differ from the nominal location of the split-gate gap (shown by a thick, white line), indicating that the domain wall is mobile during the nucleation process. Furthermore, upon annealing the sample at $\approx$ \SI{453}{K}, the domain wall position changed drastically, as shown in the right panel of Fig.~\ref{fig:DW}b, where the initially nucleated position is shown with a dashed line, and the final position, after annealing is shown with a solid line. We would like to note that the observation that a domain wall may persist even when heated above the N\'{e}el temperature is not a new one, and that this phenomenon was already explored by Brown in 1969 \cite{Brown1969a}. We also show an additional stray field image of the domain wall in Fig.~\ref{fig:DW}c, taken at the bottom right of the cyan line shown in the inset. This emphasizes the fact that the domain wall, in the absence of mesa structures and strong pinning centers, is indeed smooth at the micrometer scale and suggests that the pinning we observe is primarily due to the interaction between mesa and DW. Further examples of this pinning are shown in the bottom two panels of Fig.~\ref{fig:DW}c. In these two images, we observe simultaneous pinning at multiple mesa edges following both nucleated instances of the domain wall.  	

\section{Metropolis Hastings}\label{sec:MH}

\begin{figure}[h]
		\centering
		\includegraphics[width=86mm]{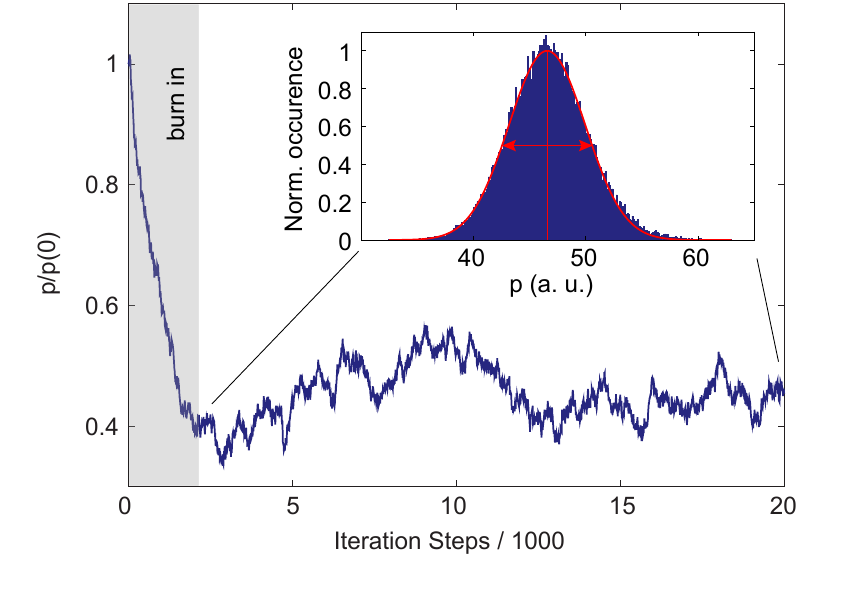}
			\caption{\textbf{Typical single parameter evolution in the Metropolis Hastings algorithm} The parameter values relative to its starting value as a function of the iteration steps of the algorithm. The initial burn-in period is shown by the shaded area. The inset shows the histogram of values taken after the burn-in period, fitted with a Gaussian distribution, with the mean and FWHM shown with red bars.}
		\label{fig:MH}
	\end{figure}

	We use a form of the Metropolis Hastings (MH) algorithm to infer the probability distributions of parameters in difficult-to-sample data sets \cite{robert2016,WileySons2012}. This MH-based parameter estimation is chosen, since the model involves correlated parameters and exhibits many local good fits to the data. Such conditions make it difficult for gradient descent methods to determine a global minimum in the difference between the data and the model as typically characterized by the mean squared error (MSE). Additionally, the analysis via the MH algorithm allows us to better estimate the uncertainty on the involved parameters by combining several datasets. For all analyses of the magnetic and sensor properties (parameters $p$) discussed in the main text, we evaluate the recorded stray field (data $D$) with a theoretical description (model) using the following implementation of this iterative algorithm (with $n$ steps):
	\begin{enumerate}
		\item A set of initial starting parameters (${p_{\textrm{curr}}}$), a step size ($d_i$) for each parameter and a theoretical model are defined.
		\item A second new candidate set of parameters $p_{\textrm{new}}$ is drawn from a proposal distribution. For this, a symmetric normal distribution is used, which is centered around the current values $p_{\textrm{curr}}$ with a width = $2d_i$, which realizes a random-walk MH algorithm to find the next candidate parameters.
		\item The model function is then computed for both parameter sets and compared with the measured data $D$ to estimate the two likelihoods $r_{\textrm{curr}}$ and $r_{\textrm{new}}$, of the data, given $p_{\textrm{curr}}$ or $p_{\textrm{new}}$, respectively. We assume unbiased Gaussian noise on the data and Jeffreys priors on the variance \cite{WileySons2012}. In particular, the likelihood is evaluated as $r_{\textrm{new/curr}} \propto (R+1)^{-(\nu+1)/2} \approx R^{-(\nu+1)/2}$, where $\nu =|D|-|p_{\textrm{new/curr}}|$ ($|\cdot|$ is the size of the set) and $R$ is the MSE of our model given the data and model parameters. Often additional prior knowledge is available on certain parameters (e.g. as a restriction to bounds or normal distributions based on previous data), in which case, we multiply these to $r$ following the Bayesian rule for the posterior. The probability for accepting $p_{\textrm{new}}$ is then realized following the typical iterative operational implementation:

		\begin{itemize}
		\item Select a random value $a$, uniformly distributed between 1 and 0.
		\item \textbf{If $(r_{\textrm{new}}/r_{\textrm{curr}})\leq a$:} Accept the new parameters 			$p_{\textrm{new}}$  and set $p_{\textrm{curr}} = p_{\textrm{new}}$.
		\item \textbf{Else:} Keep the parameter set $p_{\textrm{curr}}$.
		\item Draw a new candidate set based on $p_{\textrm{curr}}$.
		\item Repeat this procedure $n$ times.
		\end{itemize}

	\end{enumerate}
	We apply this analysis to both the mesa and domain wall stray fields using the model functions and data acquisition described in the following sections.
	\\ A typical evolution of a single parameter normalized to its starting value  is shown in Fig.~\ref{fig:MH} as a function of the iteration number. In the initial period, the parameters evolve towards the region of a better model representation of the data (higher likelihood) before it settles to a stochastic walk around a particular value (thermalization). The iteration steps before reaching this point are dropped when later examining the distribution of values, and are referred to as the "burn-in region" \cite{robert2016,WileySons2012}. The remaining steps are then processed into histograms for each parameter value, obtaining a marginalized and unnormalized probability distribution for each parameter. We test for an underlying correlation of the steps by only considering every $\eta^{\textrm{th}}$ value (thinning). In the last step, the distribution curves are approximated by Gaussians to estimate a mean and a standard deviation for a given parameter based on the data and model used.  
	
\section{Mesa Stray fields} \label{sec:mesafit}

\begin{figure*}[th]
	\centering
	\includegraphics[width=1\textwidth]{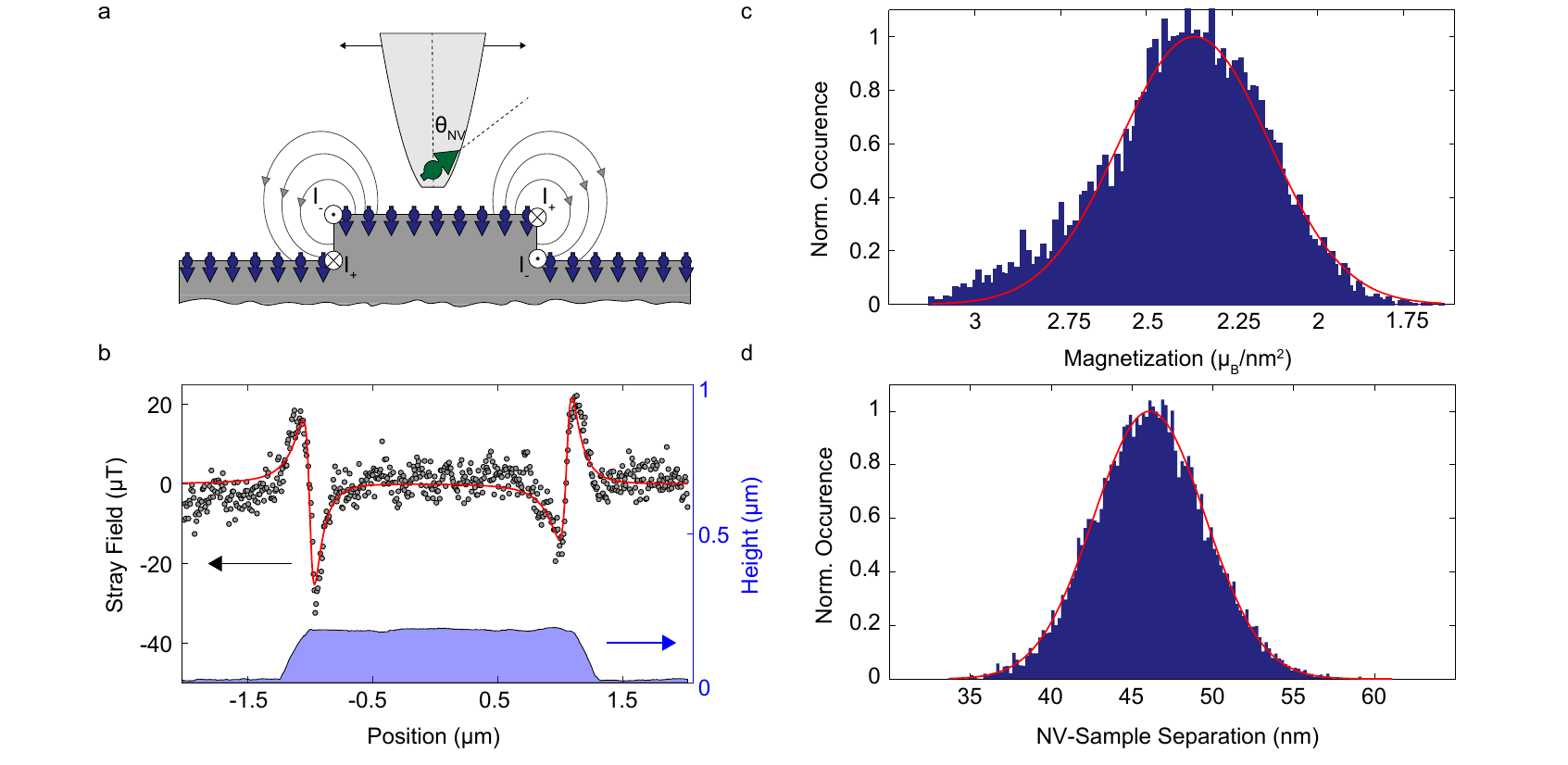}
	\caption{\textbf{Fitting of the mesa stray field} \textbf{a} Schematic of a mesa with the surface magnetization shown as an array of oriented spins. The stray field lines are shown in grey, originating from effective currents, $I_+$ and $I_-$, at the edges of the mesas. The NV is located at the tip of the scanning probe, which is scanned relative to the mesa. \textbf{b} Stray field of the mesa as measured along the NV axis with the fit of the model shown in red. The topography of the stripe is shown at the bottom. \textbf{c,d} Histograms of the results of the MH algorithm for the surface magnetization (c) and NV-to-sample spacing, $d_{NV,0}$ (d). These are fit with Gaussian distributions (shown in red) and the mean and standard deviation for each distribution is given in the text.}
	\label{fig:mesa}
	\end{figure*}

In order to extract quantitative magnetic and sensor information from a mesa, we record its stray field while scanning the magnetometer in a line that crosses the mesa (linecut). This data is then compared to a well-established model \cite{tetienne2015} for the stray field of a magnetic stripe, where the field arises from effective currents ($I_+,~I_-$) running along the top and bottom of its edges as shown in Fig.~\ref{fig:mesa}a.
According to this model, the stray field measured at a distance $d_{NV}$ from a single edge of a mesa, oriented along the y-axis, is given by:
	
         \begin{eqnarray}
	B_{NV} &=& \sin(\theta_{NV})\cos(\phi_{NV})B_x + \nonumber \\
	& & \sin(\theta_{NV})\sin(\phi_{NV})B_y + \cos(\theta_{NV})B_z,
	\label{eq:mesa}
	\end{eqnarray}
	
	\begin{eqnarray}
	B_x & = \frac{-\mu_0\sigma_m}{2\pi}\left(\frac{d_{NV}}{(x-x_0)^2 + d_{NV}^2} - \frac{(d_{NV}+t)}{(x-x_0)^2 + (d_{NV}+t)^2} \right), \\
	& ~\textrm{and} \nonumber \\
	B_z & = \frac{\mu_0\sigma_m}{2\pi}\left(\frac{x-x_0}{(x-x_0)^2 + d_{NV}^2} - \frac{x-x_0}{(x-x_0)^2 + (d_{NV}+t)^2} \right).  
	\end{eqnarray}

Here, $\sigma_m$ is the magnetization, $t$ is the thickness of the mesa, $x_0$ is the location of the edge, $\theta_{NV}$ and $\phi_{NV}$ are the polar and azimuthal angles of the NV axis respectively and $\mu_0 = 4\pi\times10^{-7}$ N/A$^2$ is the vacuum permeability. Note that $B_y = 0$ in this configuration. These equations describe the stray fields ($B_x$ and $B_z$) on one side of the mesa, and as such, we add the corresponding terms for the second edge (located at $x_1$). In order to take into account a possible asymmetric tip shape or accumulation of dirt during scanning, we allow for two different NV distances ($d_{NV}$ and $d_{NV}+\Delta d$) for either side of the mesa, as described in \cite{Rohner2019a}.
	
	With this particular analytical form, we now address the recorded stray field data and are left with seven fitting parameters ($\sigma_m,d_{NV},\theta_{NV},\phi_{NV},{x_0},{x_1}, \Delta d$).	In the first step we seek to infer the sensor orientation ($\theta_{NV}$,$\phi_{NV}$), since we can assume it to be constant throughout all measurements. For this, we perform an initial least-squared fit, seeded from 50 different initial parameter sets and choose the best fit parameters. These parameters, together with the measured stray field values and model described above (Eq.~\ref{eq:mesa}), are then used to initialize the MH algorithm (see Section \ref{sec:MH}). From all the combined datasets (29 individual linecuts), we infer the likelihood-distribution of the sensor orientation, by multiplying the individual likelihood distributions of the sensor-angles from each dataset. The resulting distribution is then described by a Gaussian, yielding a $\theta_{NV}$ and $\phi_{NV}$ of $60.7\pm2.9\deg$ and $260.6\pm0.8\deg$ respectively. 

While there is no reason for $\theta_{NV}$ and $\phi_{NV}$ to vary between scans, we can not assume the remaining parameters to stay constant throughout all datasets. Therefore, we proceed with the analysis of individual linescans, with the global sensor orientation as prior knowledge. The 29 individual parameter sets are iterated until the resulting (unnormalized) probability distributions are smooth ($n\approx 5\times10^6$ iterations) and approximate them by Gaussians. Note that here, we need to account for the position error arising from the open-loop scanner as described in Section~\ref{sec:error}. This is done by calibrating our length scale and the statements on the error of the individual parameters.

An example of this analysis for the data set in Fig.~\ref{fig:mesa}b is shown in Fig.~\ref{fig:mesa}(c,d). We approximate each of these histograms with a Gaussian and extract its mean and the standard deviation. For this particular data set, we extract $\sigma_m = 2.4 \pm 0.2$ $\mu_B$/nm$^2$, $d_{NV,0} = 46\pm 3$~nm and $d_{NV,1} := \Delta d + d_{NV,0} = 53\pm 3$~nm (not shown). In the main text, we state the mean of $\sigma_m$ obtained at room temperature together with the systematic error. 

\section{Domain Wall Model} \label{sec:DWmodel}

In order to derive a model for the stray field of a domain wall, we consider the surface magnetization of \CrO{}. As for the mesa measurements (Section \ref{sec:mesafit}), in our experiments the emerging stray field $B_{NV}$ is measured along the NV axis:
 
	\begin{equation}
	B_{NV} = B_0 + \cos(\theta_{NV})B_z + \cos(\phi_{NV})\sin(\theta_{NV})B_x,
	\end{equation}

To calculate this field, we use the procedure of propagating fields in Fourier space (with momentum vector $\textrm{q}$) \cite{Dovzhenko2018}, where for a given magnetization $M(\textrm{q})$, the Fourier components of the magnetic strayfield for a given sensor distance $d_{NV}$ can be calculated with the corresponding propagator $D(\textrm{q},d)$:

	\begin{equation}
	B(\textrm{q},d_{NV}) = D(\textrm{q},d_{NV})M(\textrm{q}),
	\end{equation} 

with:
	
	\begin{widetext}
	\begin{equation}
	D(\textrm{q},d) = \frac{\mu_0 M_s}{2}(e^{-d\textrm{q}} - e^{-(d+t_m)\textrm{q}})\begin{bmatrix}
	-\cos^2(\phi_\textrm{q}) & -\frac{\sin(2\phi_\textrm{q})}{2} & -i\cos(\phi_\textrm{q})\\ 
	-\frac{\sin(2\phi_\textrm{q})}{2} & -\sin^2(\phi_\textrm{q}) & -i\sin(\phi_\textrm{q})\\
	-i\cos(\phi_\textrm{q}) & -i\sin(\phi_\textrm{q}) & 1
	\end{bmatrix}.
	\end{equation}
	\end{widetext}
	
	Here, $t_m$ is the thickness of the magnetic layer and $M_s$ [\SI{}{\uA/\m}] is the saturation magnetization. For the two-dimensional surface magnetization, we consider the limiting case $t_m \cdot \textrm{q} \ll 1$ such that, $\sigma_m = M_s \cdot t_m$ is the surface magnetization, and the exponential pre-factor simplifies to $\frac{\mu_0\sigma_m \textrm{q}}{2}(e^{-d\textrm{q}})$.

\begin{figure*}[th]
		\centering
		\includegraphics[width=1\textwidth]{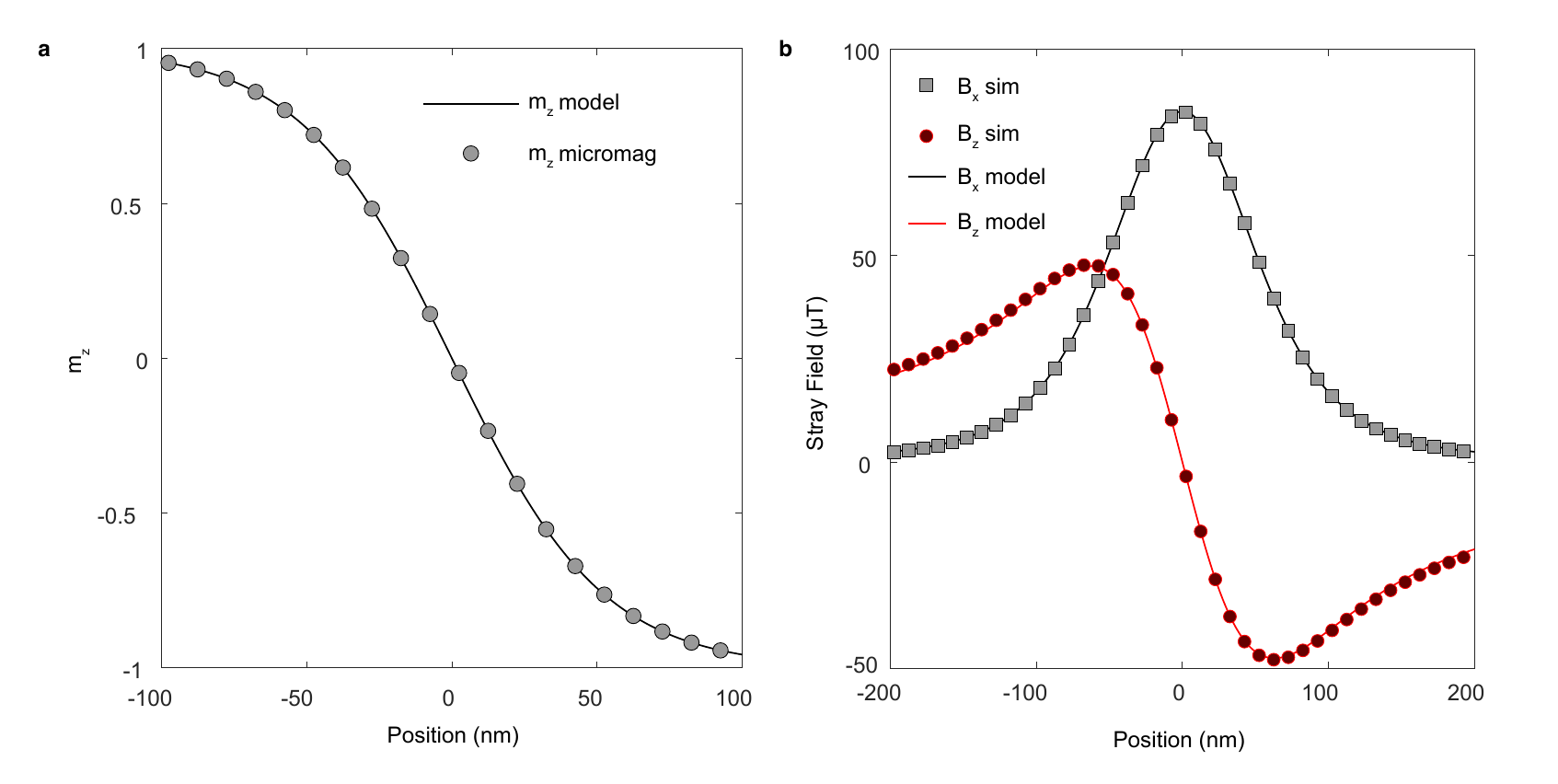}
		\caption{\textbf{a} Normalized magnetization profile of a Bloch domain wall obtained from micromagnetic simulations (black dots) and its analytical description (solid line). \textbf{b} Stray field components $B_x$, $B_z$ (black square, red dots) simulated for a \SI{20}{\nm} distance from the sample surface and calculated stray fields according to the analytical model (lines). For better visibility in both sub-panels, only every 10th data point of the simulation is shown. }
		\label{fig:DWprofile}
	\end{figure*}

After an inverse Fourier transformation, one obtains the real-space $x$ and $z$ components of the stray field of a Bloch wall:
	\begin{widetext}
	\begin{align}
	B_x & = -\frac{\mu_0\sigma_m}{2\pi^2\ell_{m}}\Re\left[-\psi^{(1)}\left(\frac{2d_{NV}+\pi \ell_{m}+2ix}{2\pi \ell_{m}}\right)+\psi^{(1)}\left(\frac{2d_{NV}+\pi \ell_{m}-2ix}{2\pi \ell_{m}}\right)\right],\\
		B_z & = \frac{\mu_0\sigma_m}{2\pi^2\ell_{m}}\Im\left[-\psi^{(1)}\left(\frac{2d_{NV}+\pi \ell_{m}+2ix}{2\pi \ell_{m}}\right)+\psi^{(1)}\left(\frac{2d_{NV}+\pi \ell_{m}-2ix}{2\pi \ell_{m}}\right)\right].
		\label{eq:DWstrayfield}
	\end{align}
	\end{widetext}	
	Here, $\psi^{(1)}$ represents the first derivative of the log gamma function. Note that the situation is translation invariant along the y-direction, and therefore produces no stray field in $B_y$.
\\
	In order to validate this analytical description of the stray field of a domain wall, we simulate a Bloch wall in MuMax3 \cite{Vansteenkiste14}. 
	In the simulation, the surface magnetization of \CrO{} is approximated by a thin slab given by a single cell with a \SI{1}{\nm} extent and a magnetization $M$ of $|M| = 10$ kA/m , exchange stiffness $\mathcal{A}=0.423$ pJ/m and uniaxial anisotropy of $K = 215.86$ J/m$^3$. 
	The total dimensions of the simulated sheet are $4096 \textrm{ nm} \times 32 \textrm{ nm} \times 1 \textrm{ nm}$ discretized to a grid of $1 \textrm{ nm} \times 2 \textrm{ nm} \times 1 \textrm{ nm}$ cells, including periodic boundary conditions to minimize boundary artifacts. 
	\\To nucleate a domain wall, we start by considering a sheet magnetized upwards in one half and downwards in the other. After energy minimization of the system via relaxation, a time-span of about \SI{1}{\us} is simulated to ensure a static equilibrium. The magnetization profile of the domain wall is then extracted and fitted according to the wall profile described by Eq. (1-3) in the Methods, which yields $\ell_m$. In the next step, the stray field at a distance of \SI{20}{\nm} is extracted from the simulation and compared with Eq.~\ref{eq:DWstrayfield}. We find very good agreement between the numerical estimates and analytical approximations. In fact, we believe the analytical description to be more accurate in capturing the dipolar stray fields, since the model considers an infinitely extended magnetic system, without the need of periodic boundary conditions or finite extent as present in the simulations. Such a comparison is shown in Fig.~\ref{fig:DWprofile}.
	
\section{Domain Wall Fitting}

Based on concurrent measurements taken over the mesa structures, we can place tight restrictions on the values of $\sigma_m$, $\theta_{NV}$, $\phi_{NV}$ and $d_{NV}$. In particular, for $d_{NV}$, we use the estimate for the larger of the two extracted distances ($d_{NV,0}$, $d_{NV,1}$, see Section~\ref{sec:mesafit}) to account for possible dirt when scanning.
The scanning direction with respect to our NV axis is readily obtained through our position calibration (Section~\ref{sec:error}), though again, taking into account the error on the angle. This leaves only the domain wall position and magnetic length $\ell_{m}$ as complete unknowns.

\begin{figure*}[th]
		\centering
	\includegraphics[width=1\textwidth]{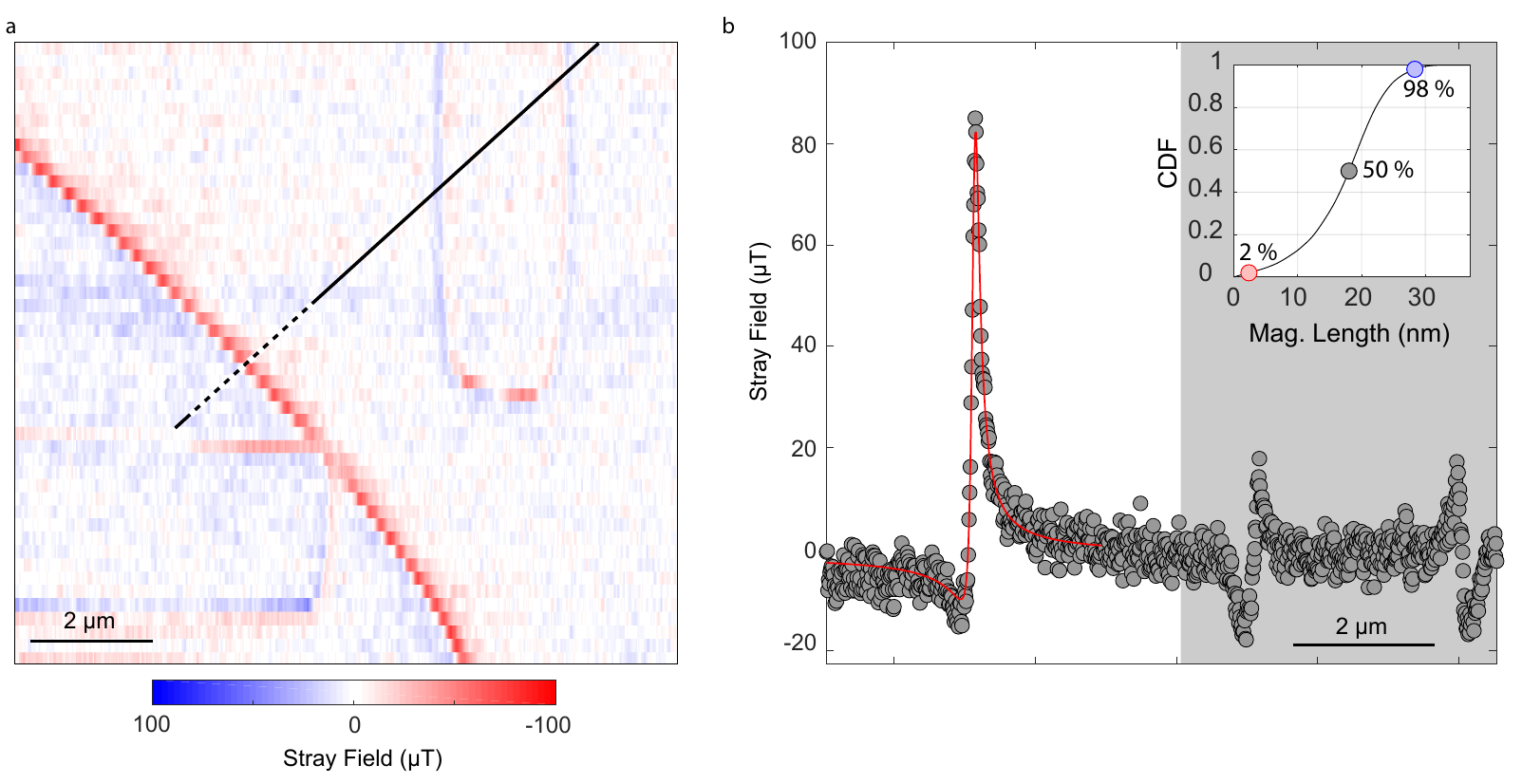}
		\caption{\textbf{Fitting of the domain wall} \textbf{a} Full field image of the domain wall running between two perpendicular mesas. The black line shows the location of the linescan, with the domain wall portion given by the dashed line. \textbf{b} The stray field over the entirety of the linecut showing the domain wall fitted with the MH algorithm (red) and the mesa stray field in the grey area. The inset shows the CDF of the distribution of magnetic lengths with the 2\%, 50\% and 98\% points shown with colored circles.}
	\label{fig:DWFit}
	\end{figure*}

To proceed, the stray field data of the domain wall, prior information and stray field model (as derived in Section~\ref{sec:DWmodel}) is analyzed via our MH algorithm implementation (Section~\ref{sec:MH}). We iterate through $n\approx20\times10^6$ steps, resulting in a reasonable modeling of the recorded stray field data, as shown in Fig.~\ref{fig:DWFit}b and smooth probability distributions for all parameters, in particular for the remaining magnetic length ($\ell_{m}$) that determines the width of the domain wall.

The inset of Fig.~\ref{fig:DWFit}b shows the cumulative distribution function (CDF) of $\ell_m$ for this data set. The measurement is taken at \SI{302}{K}, and yields a mean magnetic length of \SI{20}{\nm} with the 2$^{nd}$ and 98$^{th}$ percentiles being \SI{2}{\nm} and \SI{28}{\nm} respectively.

We claim that statements on such small $\ell_m$ ($\ell_m\ll d_{NV}$) parameters are still reasonable due to the immediate lateral and temporal proximity of the mesa and domain wall measurements. This allows us to assign any broadening in the stray field of the DW, exceeding the expected broadening from the sensor distance $d_{NV}$, to the domain wall width given by $\ell_m$. This is essentially a deconvolution of the domain wall data with the detection function of our setup, possible by the concurrently taken mesa measurement data. The analysis yields our statistical confidence in the model parameters given in the data. In our experience, most systematic errors can be excluded, with the exception of the error arising due to the piezo non-linearity, drift and hysteresis characterized in Section~\ref{sec:error}. We consider the possibilities of these errors in our statement of the DW upper bound by referring to the $98^\textrm{th}$ percentile, while remaining within a reasonable range of values. 

\section{Temperature Dependence} \label{sec:tempdep}
	By mounting the sample on a small Peltier element, we are able to access a range of sample temperatures from room temperature (\SI{295.7}{K}) up to $\approx$\SI{340}{K}. As such, we are able to explore the temperature dependence of both the sample magnetization and $\ell_m$. 
	\begin{figure*}[th]
	\centering
	\includegraphics[width=1\textwidth]{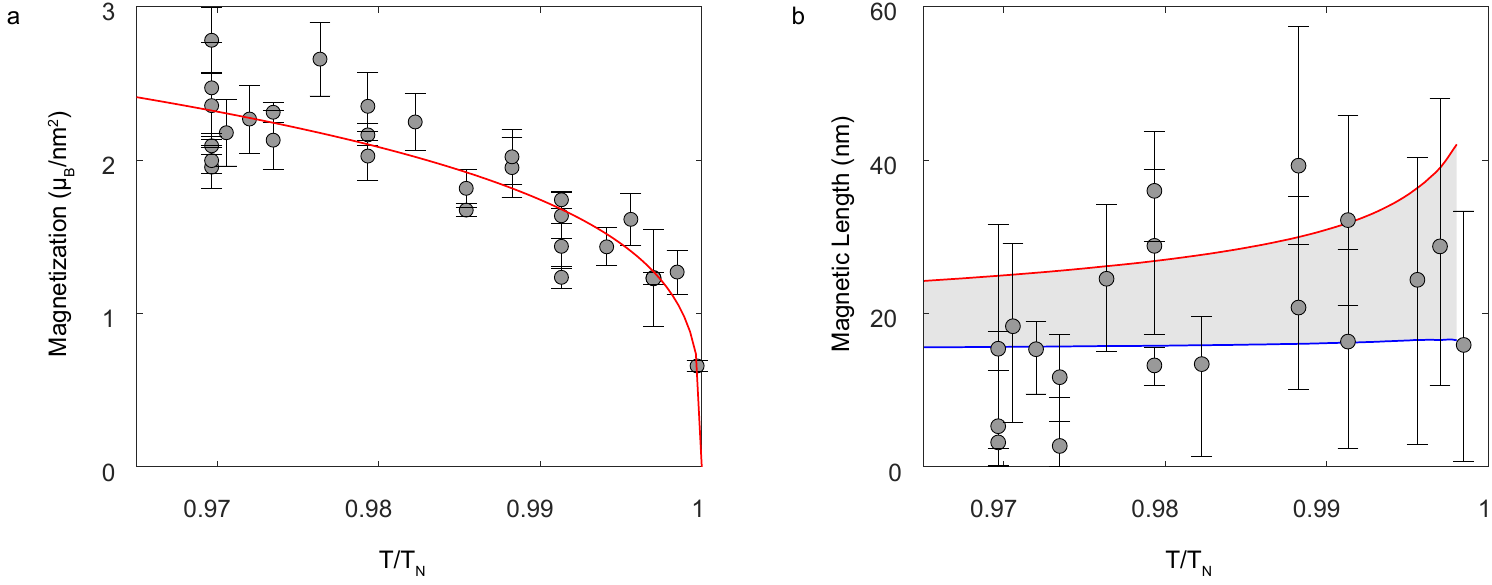}
	\caption{\textbf{Temperature dependence of $\sigma_m$ and $\ell_m$} \textbf{a} The surface magnetization, as extracted from the mesa fits, plotted as a function of temperature. The error bars in the magnetization are given by the standard deviation of the extracted magnetization distributions (see Section~\ref{sec:mesafit}). The fit is given by Eq.~\ref{eq:MsT} for $T_\textrm{N\'eel} = 307$K. \textbf{b} The magnetic length ($\ell_m$) is plotted as a function of temperature, with the mean DW parameter given by the circular points. The upper and lower limits, shown with black bars, are determined by the 98\% and 2\% confidence intervals respectively. The range of theoretical DW parameter values is presented by the shaded region between the red and blue lines. Note that all errors in the horizontal axis are smaller than the symbols.}
	\label{fig:MsVT}
	\end{figure*}
	We begin by presenting the magnetization, extracted as described in Section~\ref{sec:mesafit}, and plotted against the temperature (normalized to the N\'{e}el temperature, $T_\textrm{N\'eel}$) in Fig.~\ref{fig:MsVT}a. Here, we see that the surface magnetization falls off as 
	\begin{equation}
	\sigma_m = \sigma_{m_0}\left(1-\frac{T-T_0}{T_\textrm{N\'eel}}\right)^\beta,
	\label{eq:MsT}
	\end{equation}
	where $\beta$ is the critical exponent. Note that we allow for an offset, $T_0$ of the temperature to take into account a possible calibration offset between the thermistor (which defines $T$) and the actual sample temperature. Here, we assume the literature value of the N\'{e}el temperature (\SI{307}{K}), a reasonable assumption in the absence of excessive strain~\cite{Kota13} or doping~\cite{Mu13}, which may lead to changes in $T_\textrm{N\'eel}$.  In doing so, we obtain a relatively small offset of only \SI{2}{K} and a critical exponent $\beta = 0.26$, well within the range of previously measured values \cite{mahdawi2017}. 
	
	Note that the vertical error bars in Fig.~\ref{fig:MsVT}a are given by the standard deviation of a Gaussian fit to the distribution of $\sigma_m$, extracted using the MH algorithm. 
	We now repeat the same procedure for $\ell_m$, plotted against the temperature in Fig.~\ref{fig:MsVT}b. Here, the filled circles represent the mean magnetic length extracted from the MH algorithm for each data set. The error bars are given by the 98$^{th}$ (2$^{nd}$) percentiles of the CDF, to show the maximum (minimum) values of $\ell_m$ that would be consistent with our data given the constraints we place based on the mesa measurements. In particular, we can look at the room temperature measurements (first three points at the left), which all fall below the 98$^{th}$ percentile bar at \SI{32}{\nm}. This justifies our statement in the main text, that $\ell_m>32$ nm is inconsistent with our measurements at room temperature. Furthermore, we show the theoretical upper (blue) and lower (red) limits given by $\ell_m = \sqrt{\mathcal{A}/\mathcal{K}}$ where $\mathcal{A}$ is the exchange stiffness and $\mathcal{K}$ is the anisotropy~\cite{Foner1963a}. In particular, we use the following temperature dependence for $\mathcal{A}$~\cite{Koebler12,Parthasarathy2019a}:
	\begin{equation}
	\mathcal{A}(T) = \mathcal{A}(0)\left[\frac{m(T)}{m(0)}\right]^\alpha,
	\end{equation}
	where $m$ is the sublattice magnetization and $\alpha\in[1,2]$. The exact value of $\alpha$ is unknown, but is believed to be close to $\alpha = 2$ yielding generally smaller domain wall widths. Furthermore, the value of $\mathcal{A}(0)$ is estimated as:
	\begin{equation}\label{eq:A0-estimation}
	\mathcal{A}(0) = \dfrac{\myJ S^2}{a},
	\end{equation}
	with the exchange integral $\myJ = 2.34\times 10^{-21}$\,J taken from DFT data~\cite{Shi09} and $S = 1$ being the effective spin length~\cite{Koebler12}. As such, we see that our measurements are consistent with the theoretical expectations, and expect that, close to the N\'{e}el temperature, we would reach a regime where $\ell_m>d_{NV}$, in which case, we could directly measure $\ell_m$. This would require either higher spatial resolution (for lower temperatures) or a higher sensitivity (close to $T_\textrm{N\'eel}$), which should be achievable in future work. 

\section{Analytics for Snell's Law}
We consider a semi-infinite sample with a mesa of width $w$ and thickness $t$ on the top surface ($z = 0$). It is assumed that the mesa has a constant rectangular cross-section and is directed along the $\bm{e}_y$ axis. The continuum model of \CrO{} can be represented using two antiferromagnetically coupled sublattices with unit magnetization vectors $\bm{M}_a(\bm{r})$ and $\bm{M}_b(\bm{r})$~\cite{Samuelsen70,Mu19}. Within the long-wave approximation, it is reasonable to use the N\'{e}el vector order parameter $\bm{L}(\bm{r}) = (\bm{M}_a - \bm{M}_b)/2$ and the total magnetization vector $\bm{M}(\bm{r}) = (\bm{M}_a + \bm{M}_b)/2$, with $|\bm{L}| = 1$  and $|\bm{M}| \approx 0$. The latter will be neglected in the following. Then, the effective energy of the sample reads~\cite{Auerbach94,Ivanov05a}
\begin{equation}\label{eq:en-eff}
\mathcal{E} = \mathcal{K} \int \left[\ell_m^2 \sum_{\nu=x,y,z} (\partial_\nu \bm{L})^2 + (1-L_z^2) \right] \mathrm{d}\bm{r},
\end{equation}
where $\ell_m = \sqrt{\mathcal{A}/\mathcal{K}}$ is the magnetic length as for the spin-lattice model. Note that additional magnetoelastic terms may be incorporated into the effective value of the anisotropy constant $\mathcal{K}$, leading to a shift in $\ell_m$ with qualitatively identical results~\cite{Dudko71,Gomonay07a,Kota16}. We furthermore assume the exchange-driven Neumann boundary conditions for the N\'{e}el vector:
\begin{equation}\label{eq:bc}
\bm{L}\times (\mathrm{\mathbf{n}}_\textsc{s}\cdot \nabla)\bm{L}= 0,
\end{equation}
where $\mathrm{\mathbf{n}}_\textsc{s}$ is the surface normal. In the following, we use the local spherical reference frame parametrization $\bm{L} = \{ \sin\vartheta\cos\varphi, \sin\vartheta\sin\varphi, \cos\vartheta\}$. We set the equilibrium, bulk domain wall position to the plane $y=kx$ where $k$ is assumed to be small. We also assume a mesa geometry satisfying $t/w > 0.01$. Then, we can describe the domain wall as it passes through the mesa through the following Ansatz:
\begin{equation}\label{eq:ansatz}
\vartheta = \begin{cases}
2\arctan\exp \frac{y' - y_0^\text{b}(x',z)}{\ell_m}, & z < 0 \\
2\arctan\exp \frac{y - y_0^\text{m}(x,z)}{\ell_m}, & z \ge 0
\end{cases}\quad \varphi = \mathrm{const}.
\end{equation}
Here, $(x',y') = R_{\bm{e}_z}(\nu)(x,y)$ with $R_{\bm{e}_z}(\nu)$ being the rotation matrix around $\bm{e}_z$ at an angle $\nu$, and $y_0^\text{b,m}$ describes the domain wall profile in bulk and mesa, respectively. We use
\begin{equation}\label{eq:y0-bulk}
y_0^\text{b}(x,z) = (k_0 - k)b \sech \dfrac{x}{b} \tanh \dfrac{x}{b} e^{-\frac{z^2}{2c^2}}
\end{equation} 
with $b =w/(2\arcsinh 1)$. The values of $k_0$ and $c$ are determined through the energy minimization below. The function $y_0^\text{m}(x,z)$ is determined by minimizing the energy functional~\eqref{eq:en-eff} within the mesa and reads
\begin{equation}\label{eq:y0-mesa}
y_0^\text{m}(x,z) = \dfrac{4k_0}{w}\sum_{n = 0}^\infty \dfrac{(-1)^{n}}{\lambda_n^2} \sech\lambda_nt \cosh \lambda_n(t-z) \sin \lambda_n x,
\end{equation}
where $\lambda_n = \dfrac{(1+2n)\pi}{w}$ and we set $y_0^\text{m}(x,0) = k_0x$. Then, the total energy of the domain wall ($\mathcal{E}$), up to a constant, reads:
\begin{widetext}
\begin{equation}\label{eq:etot}
\begin{aligned}
\mathcal{E} & = \mathcal{E}_0 k^2w^2 \left[ 1 - f\left(\frac{t}{w}\right)\right],\quad f(x) = \left[ C_2 + C_1\sum_{n = 1}^{N_0} \dfrac{ \tanh \pi (2n-1) x - 1}{(2n-1)^3} \right]^{-1},\\
\mathcal{E}_0 & = \dfrac{16\mathcal{K}\ell_m}{\pi^3 C_1},\quad C_1 = \dfrac{48\sqrt{70}\arcsinh^21}{7\pi^{7/2}} \approx 0.811,\quad C_2 = 1 + \dfrac{7}{8}\zeta(3)C_1\approx 1.853,
\end{aligned}
\end{equation}
\end{widetext}
where $\zeta(3) \approx 1.202$ is the value of the Riemann zeta-function and $N_0$ is chosen from condition $\tanh \pi (2N_0-1) x \approx 1$. The value of $c$ from Eq.~\eqref{eq:y0-bulk} is thereby given by $c = \sqrt{\dfrac{5}{14}}\dfrac{w}{2\arcsinh 1}$. This parameter plays a particularly important role as it determines the length scale over which inhomogeneities in the domain wall persist into the bulk below the mesa. For this reason, we rename this parameter as $t_B$ in the main text. Furthermore, $k_0 = kf(t/w)$ characterizes the direction of the domain wall at the bulk-mesa interface. Note, that it is dependent only on the mesa aspect ratio, which allows us to scale simulations for direct comparison with experiment. 
\begin{figure*}[th]
	\includegraphics[width=\linewidth]{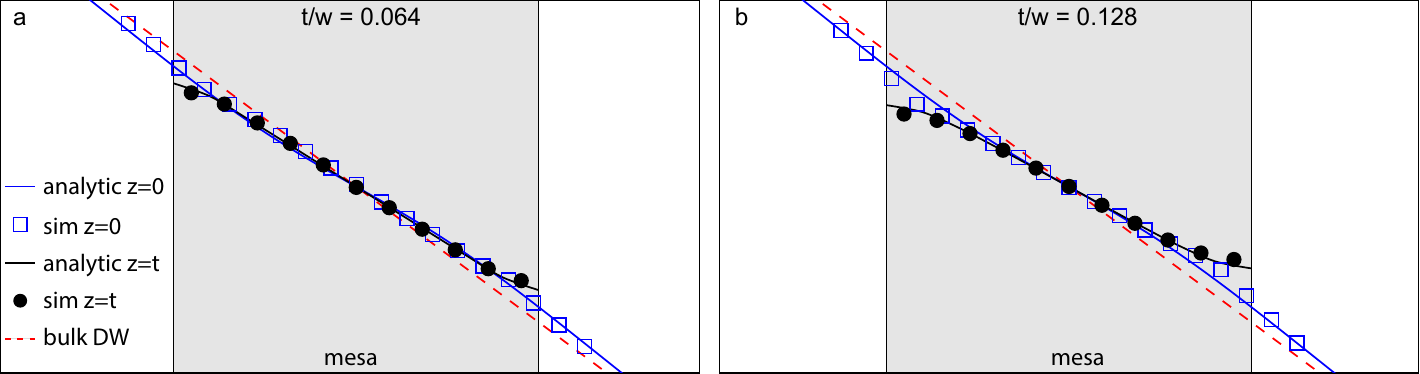}
	\caption{\textbf{Analytical analysis of the domain wall in the bulk and mesa} Comparison of the domain wall profile in analytics (solid lines) and simulations (symbols) for two mesa aspect ratios. Level $z = t$ (top of mesa, see Eq.~\eqref{eq:y0-mesa}) and level $z = 0$ (bulk-mesa interface, see Eq.~\eqref{eq:y0-bulk}) are shown by black and blue. Equilibrium direction of the domain wall in bulk is shown by red dashed line. Mesa region is colored by gray. Simulation parameters: $w = 47a$, $t = 3a$  \textbf{a} and $t = 6a$  \textbf{b}, sample dimensions (without mesa) $199a\times 199a\times 49a$.}
	\label{fig:dw-profile}
\end{figure*}
In particular, Fig.~\ref{fig:dw-profile} shows the comparison between the analytics developed here (solid black and blue lines) and simulations (black circles and blue squares, where we see excellent agreement. In both analytics and simulation, we see the S-shaped bending of the domain wall on the mesa and the gradual twisting of the domain wall to match that of the bulk position (red dashed line) as we go into the bulk. The S-shaped deviation is much less pronounced for thinner mesas (Fig.~\ref{fig:dw-profile}a), which is very similar to the experimental case.

The Snell's law for the domain wall can be determined using the equilibrium domain wall profile in bulk and at the mesa top surface ($z = t$). The incidence angle is given by $\theta_\text{1} = \arctan k$, while the refraction angle can be estimated as $\theta_\text{2} = \arctan k_1$ with $k_1 := \partial_x y_0^\text{m}(0,t)$. Then,
\begin{equation}\label{eq:snell}
\dfrac{\sin\theta_\text{1}}{\sin\theta_\text{2}} = \dfrac{k}{k_1} \sqrt{\dfrac{1+k_1^2}{1+k^2}}\stackrel{t/w\to 0}{\approx} 1 + 3.1\cos^2\theta_1 \dfrac{t}{w}.
\end{equation}

\section{Elastic Properties of the Domain Wall}
\begin{figure*}[th]
	\centering
	\includegraphics[width=0.8\textwidth]{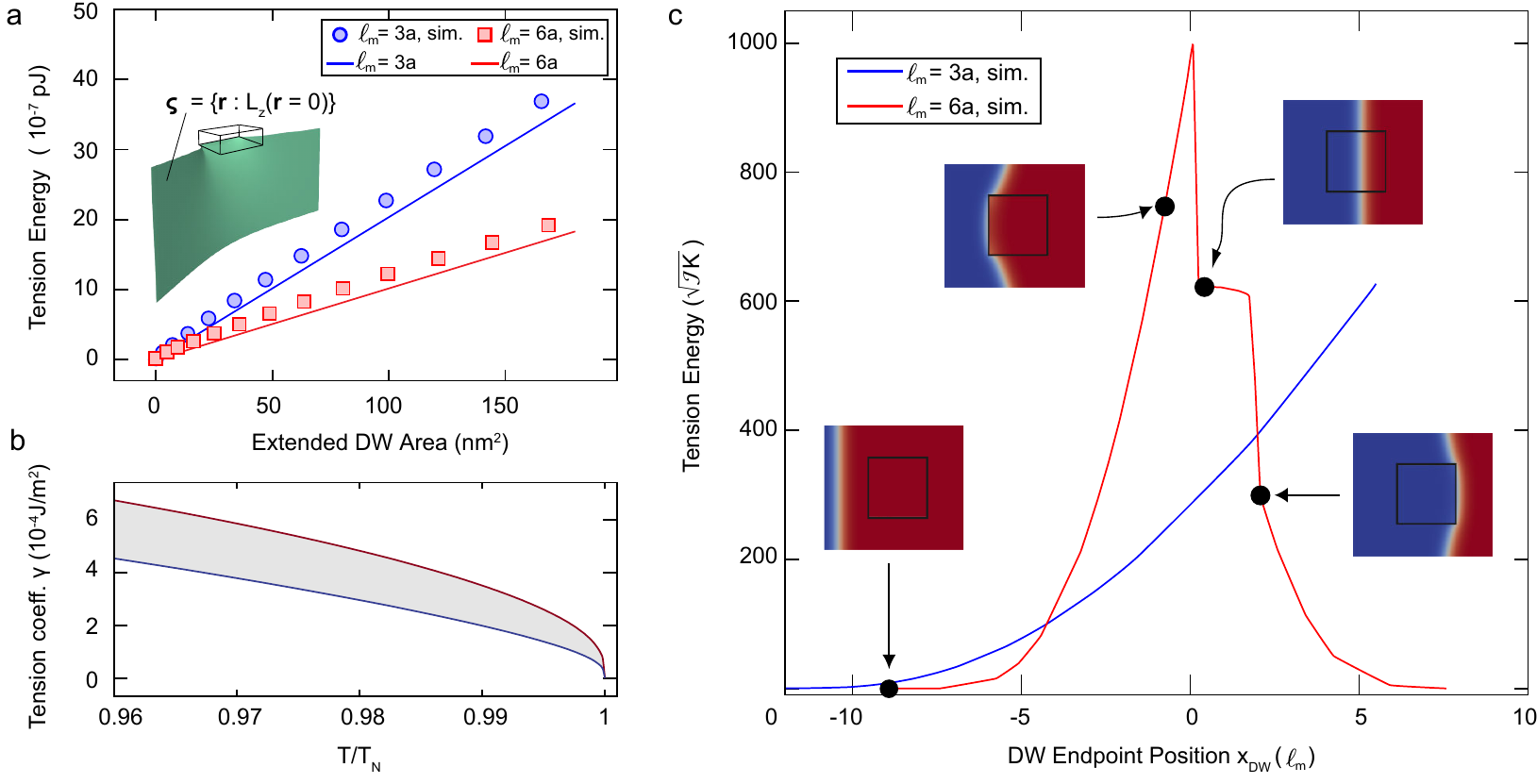}
	\caption{\textbf{Extended simulations on DW elasticity} \textbf{a} Surface energy of the DW as a function of the increased DW area arising due to a bend around a mesa. We compare simulations (circles and squares) and calculations (lines) for two different magnetic length values $\ell_m$ = 3a (blue) and $\ell_m$ = 6a (red) \textbf{b} Temperature dependence of $\gamma$ showing upper and lower bounds determined as in Fig. S7, showing the softening of the DW elasticity with increasing temperatures. \textbf{c} Pinning behavior of the DW for two effective temperatures (set by changing the magnetic length), where the blue curve ($\ell_m$ = 3a) is at a lower temperature than the red ($\ell_m$ = 6a). The insets show snapshots of the simulated DW position along the red curve. The blue curve corresponds to that seen in Fig.~3c in the main text.}
	\label{fig:tension}
\end{figure*}

In experiments and simulations, we have observed a behavior of the DW that mimics that of a rubber band. As such, we describe the DW trajectory and interactions with the mesa using its elastic properties, that is, the DW surface energy and corresponding surface tension. In particular, we can consider the surface $\bm{\varsigma}$ where the N\'{e}el order parameter lies horizontally ($L_z(\bm{r}) = 0$), and use this to describe the DW. We can address the details of $\bm{\varsigma}$ by means of spin lattice simulations (see Simulation Details in the main text for more details). In Fig.~2c of the main text, we see that the domain experiences the strongest deflection at the edge of the mesa when crossing from bulk to mesa. As previously discussed, this behavior is well-described by the Ansatz in Eq.~\ref{eq:ansatz}, where the $b$ parameter determines how far the wall is deflected in the plane of the mesa and $c$ (introduced in Eq.~\ref{eq:y0-bulk}) characterizes the deflection in the vertical direction. The same behavior exists in the case where the domain wall is deformed around the mesa, as shown in the inset of Fig.~\ref{fig:tension}a. Here, $\bm{\varsigma}$ exhibits a smooth bend deep in the bulk and sharper deflections near the mesa edges. The additional, tensional energy due to this bending is plotted here for two different values of $\ell_m$ (circles and squares) as a function of the increase in area DW area $\mathcal{S}$ arising from the bending. We can compare the results of these simulations with analytical calculations if we assume that the inhomogeneities of $\bm{\varsigma}$ are gradual, i.e. have a radius of curvature larger than $\ell_m$. In this case, we can describe the DW's mechanical tension by:
\begin{equation}
\gamma = \mathcal{E}/\mathcal{S} = 4\sqrt{\mathcal{A}\mathcal{K}}.
\end{equation}
Here, we have defined a tension coefficient $\gamma$ for the DW, which is plotted in Fig.~\ref{fig:tension}a as solid lines for two different values of $\ell_m$. 

We can furthermore explore the impact of temperature on the tension coefficient. To do so, we use the temperature dependence of $\mathcal{A}$~\cite{Koebler12,Parthasarathy2019a} and $\mathcal{K}$~\cite{Foner1963a}, as in Fig.~\ref{fig:MsVT}b. The corresponding upper and lower bounds of the tension coefficient are then shown in as shown in Fig.~\ref{fig:tension}b, where the observed reduction in $\gamma$ with increasing temperature implies an increase in elasticity, or a 'DW softening'. The consequences of this are more clear in Fig.~\ref{fig:tension}c. Here, we show the pinning surface (tension) energy of the domain wall as a function of the DW position relative to a mesa, as in Fig.~3c in the main text, for two different values of $\ell_m$. As $\ell_m$ increases with increasing temperature, this acts as an effective tuning parameter for the temperature in these simulations. The blue curve is as shown in Fig.~3c, where the DW deforms continuously. However, for larger $\ell_m$ (red), i.e. higher temperatures, we see a different pinning behavior, illustrated with the snapshots inset to the figure. Here, the DW is first deformed around the mesa, before snapping into a straight configuration, passing through the mesa. As the DW is moved closer to the opposite edge of the mesa, the mesa will again be preferentially pinned to the mesa edge for some distance, on the order of $\ell_m$, at which point the DW no longer feels the influence of the mesa. Thus, we expect that the pinning of the DW can be strongly influenced by temperature. However, this reduction in pinning strength can be overcome by carefully tuning the mesa geometry. 

\begin{figure*}[th]
	\centering
	\includegraphics[width=0.8\textwidth]{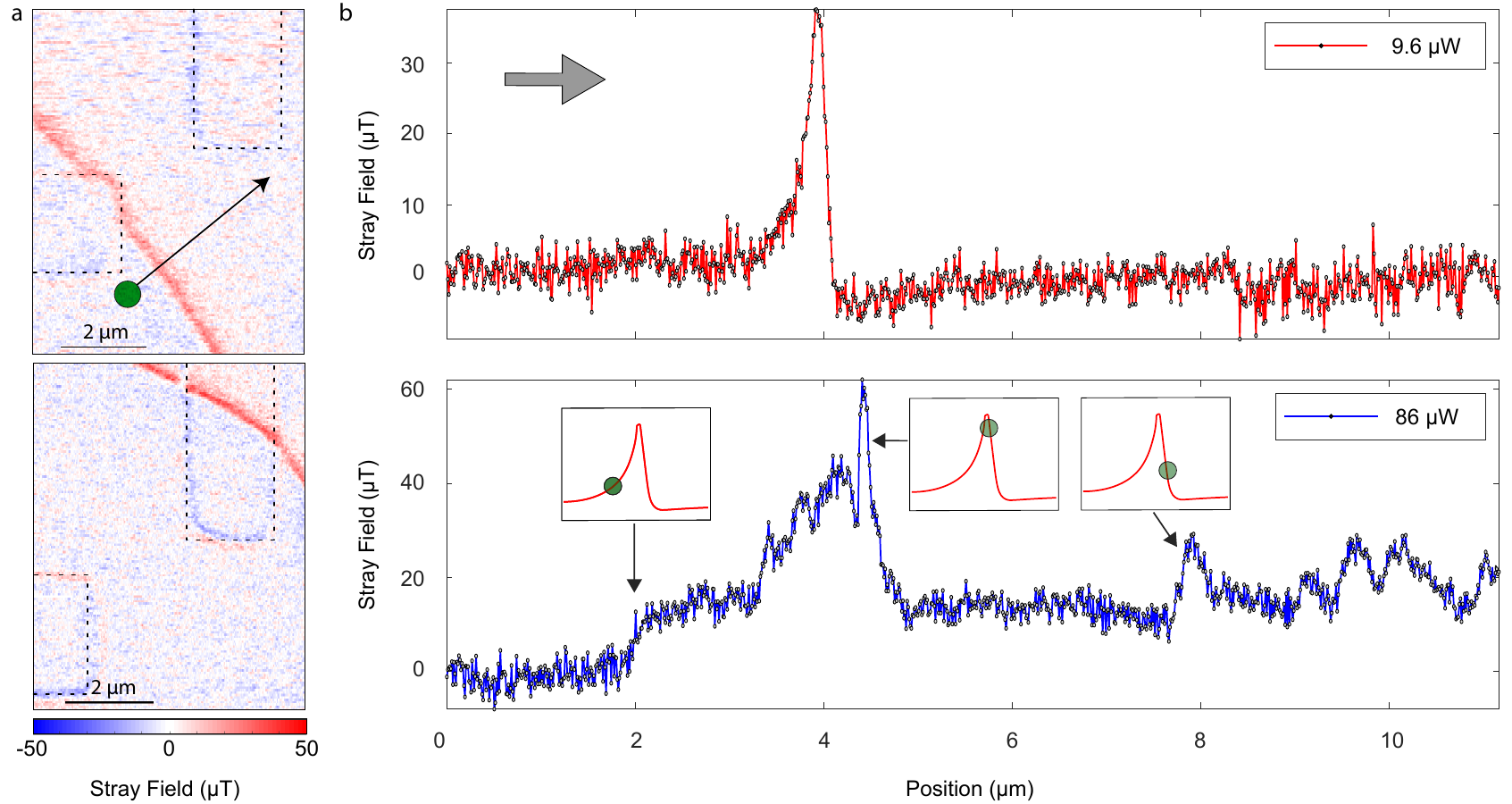}
	\caption{\textbf{Laser dragging} \textbf{a} Laser dragging achieved over a large distance. The DW, originally pinned to the corner of one mesa (top) is dragged via laser in the direction shown by the black arrow. A second image (below) shows the DW becoming pinned on top of an adjacent mesa. \textbf{b} Two linescans showing NV magnetometry at low power (upper curve, red) and high power (lower curve, blue), with the powers noted in the upper right corners. The grey arrow in the top corner shows the scanning direction. In the top image, we scan across the domain wall without disturbing it, while in the bottom, we form an attractive potential, causing the domain wall to move. The insets show the relative position of the NV (green circle) to the domain wall stray field (red curve) at several positions.}
	\label{fig:drag}
\end{figure*}

\section{Domain Wall Dragging}
We are able to explain dragging of the domain wall by the laser through the formation of an effective attractive potential for the DW by local heating. Heating of the sample appears to lower the effective depth of pinning potentials, evidenced by the fact that reproducible laser dragging is only achieved at sample temperatures above room temperature, and near $T_\textrm{N\'eel}$. The additional heating provided by the laser then allows us to completely overcome this barrier, and move the domain wall freely. However, once the heating, i.e. laser, is removed, the domain wall tension causes the wall to snap back to its original position unless it becomes pinned along the way. This dragging and pinning can also be achieved over $\mu$m-scale distances, as shown in Fig.~\ref{fig:drag}a, where we drag the DW using the same technique outlined in the main text, over a distance of \SI{6}{\um}. This implies that every time we observe a movement of the DW, we are moving it between strong pinning sites, which can be internal crystal defects or fabricated surface structures. 
\\We support these claims by further scans across the domain wall with the NV scanning probe at increased laser intensities. In this way, we aim to heat the sample with the near-field of the excitation laser near the tip of the scanning probe while simultaneously measuring the stray field from the DW at the NV position. An example of this procedure is shown in Fig.~\ref{fig:drag}b, where we compare the measured stray field at low power (\SI{9.7}{\uW} - red) and high power (\SI{85}{\uW} - blue) \SI{532}{\nm} laser excitation (with powers measured at the rear lens of the microscope objective). Each scan is performed with a \SI{3}{s} integration time per point, over the same section of domain wall, at a global sample temperature of 304.5 K. At low power, we see a domain wall stray field as already discussed. Increasing the power results in additional peaks in $B_{NV}$ occurring at erratic locations. We explain this observation with a number of pinning sites located along the DW path as the DW is bent by laser dragging. In particular, we see an initial increase and plateau in $B_{NV}$, which is consistent with an attractive potential moving the domain wall ahead of the NV position. The field then increases, meaning that we pass over the domain wall with the NV, which can only be achieved by pinning of the domain wall. However, at some point, the pinning is overcome and the domain wall is again dragged together with the laser. We furthermore see a number of smaller pinning centers, evidenced by the slight peaks in field towards the end of the scan. In the insets of Fig.~\ref{fig:drag}b, we show the position of the NV relative to the stray field pattern that would result in such peaks. Furthermore, as we continue to observe a non-zero stray field while scanning, we expect that the sample temperature surrounding the NV is still below $T_\textrm{N\'eel}$. If we again image the domain wall at low green laser power after such dragging, we see that the position remains unchanged in most cases. This indicates that the other pinning centers we observed are rather weak, and are overcome by the domain wall tension. Thus, we can directly observe the dragging of the domain wall by the laser and thereby gain information about the pinning landscape in the sample - a potential avenue for future research.

\end{document}